\documentstyle[amsfonts,amssymb,graphicx,mathrsfs,color,ulem]{article}
\newcommand{\beq}{\begin{eqnarray}}
\newcommand{\eeq}{\end{eqnarray}}
\newcommand{\beqst}{\begin{eqnarray*}}
\newcommand{\eeqst}{\end{eqnarray*}}

\newcommand{\qed}{\hfill $\square$}

\newcommand{\dsp}{\displaystyle}

\newtheorem{theorem}{Theorem}[section]
\newtheorem{lemma}[theorem]{Lemma}

\newtheorem{proposition}{Proposition}[section]

\newtheorem{remark}[theorem]{Remark}

\title{\bf 
 Fundamental solutions for the Dirac equation   in  curved   spacetime  and generalized Euler-Poisson-Darboux equation
}

\author{{\bf Karen Yagdjian${}^{\scriptsize \mbox{\rm \scriptsize 1}}$  and  Anahit Galstian${}^{\small \mbox{\rm \scriptsize 2}}$} }

\begin{document}

\date{}

\maketitle

\thispagestyle{empty}
\begin{center}
{\small {School of Mathematical and Statistical Sciences,\\
University of Texas RGV, 
1201 W.~University Drive,  \\
Edinburg, TX 78539,
USA \\
${}^1$e-mail: karen.yagdjian@utrgv.edu\\
${}^2$e-mail: anahit.galstyan@utrgv.edu}}
\end{center}
\medskip
\vspace*{-0.6cm}

\begin{abstract}
We present  the fundamental solutions for the  spin-1/2 fields propagating in  spacetimes with power type expansion/contraction  and   the fundamental solution   of the Cauchy problem  for the Dirac  equation. The derivation of these fundamental solutions  is based on formulas  for the solutions  to the generalized  Euler-Poisson-Darboux equation, which are obtained by the integral  transform approach.

\medskip

\noindent
{\bf Keywords\,\, } Dirac equation $ \cdot $   Einstein~\&~de~Sitter model   $ \cdot $ FLRW   models  $ \cdot $   non-Fuchsian equations $ \cdot $  Euler-Poisson-Darboux equation
\end{abstract}

\section{Introduction. Main results}

The Dirac equation is one of the important equations of relativistic quantum mechanics,  quantum electrodynamics, and quantum field theory. The Dirac equation and its quantization in curved spacetime are of great
interest due to the role  of spin-$\frac{1}{2}$ particles in astrophysics and cosmology. Recent observational confirmation of the expansion of the  universe and  the quantum field theory  demand a detailed investigation of the solutions of the  Dirac equation in  curved spacetime (see, e.g., \cite{Birrell,Parker} and bibliography therein). 
The construction of a quantum field theory in curved spacetime relies heavily  on a detailed description of the  solutions of the Dirac equation in
a  curved background. The standard models of Cosmology provide such backgrounds,  which form a family of curved backgrounds under the name  Friedmann-Lema\^itre-Robertson-Walker (FLRW)  models. 
For the  FLRW  spacetime with accelerated  expansion/contraction, more exactly, for the de~Sitter spacetime in \cite{AP2020} a fundamental solution of the Dirac operator and an explicit formula for the solution of the Cauchy problem are  obtained. In \cite{ArX2020} an examination  of these  explicit formulas gave the answer to such an interesting question in the physics of fundamental particles  as a validity of the Huygens principle. It must be noted that several   exact solutions of the Dirac equation in  the FLRW   spacetime, which is the basis of the standard cosmological model, 
obtained by the separation of variables approach answer to some very 
interesting questions of physics (see, e.g., \cite{Barut-D,Brill-Wheeler,Finster,Huang_Justin,Huang,Huang2005,Oliveira,Villalba,Zecca} and bibliography therein). On the other hand, the fundamental solutions to the Dirac operator 
can provide   information that is impossible to obtain by numerical solutions of differential equations, as well as new solutions that are not among  the exact solutions obtained by the separation of variables approach. 

In the present paper we construct the fundamental solution of the Dirac operator  and obtain the explicit formulas for the solution of the Cauchy problem  for the FLRW   spacetime with both accelerating and decelerating expansion or contraction. The spatially flat FLRW   models considered in this paper have 
the metric tensor that, in Cartesian coordinates, is written as follows 
\begin{equation}
\label{mtg}
(g_{\mu \nu })=    \left (
   \begin{array}{ccccc}
 1& 0& 0   & 0 \\
   0& -a^2(t) &  0 & 0 \\ 
 0 & 0 &  -a^2(t)   & 0 \\
 0& 0& 0   &  -a^2(t) \\
   \end{array} \right),\quad \mu ,\nu =0,1,2,3,  
\end{equation}
where the scale factor    $a(t)=a_0t^{\ell}  $, $\ell \in {\mathbb R}$,  $ t>0$,  and $x \in {\mathbb R}^3$, $x_0=t $. If $\ell<0$ the spacetime is contracting. In the case of $\ell>1$ the expansion is accelerating (with horizon), while for $0<\ell<1$ the expansion is decelerating. In the case of the Milne spacetime \cite{Sean Carroll,Gron-Hervik, Schrodinger}   $\ell=1 $. 
The FLRW  spacetime with the scale factors   $a(t)=a_0t^{2/3}  $ and  $a(t)=a_0t^{1/2}  $ (see, e.g., \cite{Moller,Ohanian-Ruffini})  are modeling the matter dominated universe (Einstein~\&~de~Sitter model \cite{Einstein-Sitter}) and the radiation dominated universe, respectively.  The Dirac equation in the   spacetime with the metric tensor (\ref{mtg}) is (see, e.g., \cite{Barut-D})
\begin{equation}
\label{DE}
  \dsp 
\left(  i {\gamma }^0    \partial_t   +i \frac{1}{a(t)}{\gamma }^1  \partial_{x_1}+i  \frac{1}{a(t)}{\gamma }^2 \partial_{x_2}+i \frac{1}{a(t)}{\gamma }^ 3   \partial_{x_3} +i \frac{3\dot a(t)}{2a(t)}    {\gamma }^0     -m{\mathbb I}_4 \right)\Psi=F \,,
\end{equation}
where 
the contravariant gamma matrices are   (see,  e.g., \cite[p. 61]{B-Sh})
\begin{eqnarray*}
&  &
 \gamma ^0= \left (
   \begin{array}{ccccc}
   {\mathbb I}_2& {\mathbb O}_2   \\
   {\mathbb O}_2& -{\mathbb I}_2   \\ 
   \end{array}
   \right),\quad 
\gamma ^k= \left (
   \begin{array}{ccccc}
  {\mathbb O}_2& \sigma ^k   \\
  -\sigma ^k &  {\mathbb O}_2  \\  
   \end{array}
   \right),\quad k=1,2,3\,.
 \end{eqnarray*}
Here $\sigma ^k $ are the Pauli matrices 
\begin{eqnarray*}
&  &
\sigma ^1= \left (
   \begin{array}{ccccc}
  0& 1   \\
  1& 0  \\  
   \end{array}
   \right), \quad
\sigma ^2= \left (
   \begin{array}{ccccc}
  0& -i   \\
  i& 0  \\  
   \end{array}
   \right),\quad
\sigma ^3= \left (
   \begin{array}{ccccc}
  1& 0   \\
  0&-1 \\  
   \end{array}
   \right)\,,
\end{eqnarray*}
and  ${\mathbb I}_n $, ${\mathbb O}_n $ denote the $n\times n$ identity and zero matrices, respectively. 

We  consider the fundamental solutions and the Cauchy problem with the initial time $t=\varepsilon>0 $. Furthermore,  we admit  that the mass of the field (particle)  can be changing in time and  vanishing at infinity. More exactly, the model that we choose is determined by the Dirac operator
\begin{equation}
\label{DO}
{\mathscr{D}}(t,\partial _t,\partial _x)
 := 
  \dsp 
 i {\gamma }^0    \partial_t   +i \frac{1}{a(t)}{\gamma }^1  \partial_{x_1}+i \frac{1}{a(t)}{\gamma }^2 \partial_{x_2}+i \frac{1}{a(t)}{\gamma }^ 3   \partial_{x_3} +i \frac{3\dot a(t)}{2a(t)}    {\gamma }^0     -m t^{-1}{\mathbb I}_4  \,,
\end{equation}
where $m \in {\mathbb C}$. This model includes the equation of a neutrino with $m=0$ as well as a massive spin-$\frac{1}{2}$ particle  undergoing a redshifting of its wavelength as the universe expands.

We start with the  fundamental solution to the Dirac operator. 
 Recall  that   a retarded fundamental solution (a retarded inverse) for the Dirac operator (\ref{DO})  is a four-dimensional matrix with the 
operator-valued  entries $ 
{\mathcal E}^{ret}={\mathcal E}^{ret} \left(x, t ; x_{0}, t_{0};m\right)
 $      that solves the equation
\begin{equation}
\label{FSE} 
{\mathscr{D}}(t,\partial _t,\partial _x){\mathcal E}  \left(x, t ; x_{0}, t_{0};m\right) 
 = 
\delta ( x-x_0) \delta (t-t_0) {\mathbb I}_4, \qquad 
 (x,t), (x_0,t_0 ) \in {\mathbb R}^3\times {\mathbb R}_+,  
\end{equation}
and with the support in the {\it chronological future} (causal future) $D_+(x_0, t_0)$  of the point $(x_0,t_0)  \in {\mathbb R}^3\times {\mathbb R}_+$. The
advanced fundamental solution (propagator) $ 
{\mathcal E}^{adv}={\mathcal E}^{adv}  (x, t ; x_{0}, t_{0};$ $  m )
 $ solves the equation (\ref{FSE}) and has a  support in the {\it chronological past} (causal
past) $D_-(x_0, t_0)$.  The forward and backward light cones are defined as the boundaries of  
\[ 
D_{\pm}\left(x_{0}, t_{0}\right) :=\left\{(x, t) \in {\mathbb R}^3\times {\mathbb R}_+ ;
\left|x-x_{0}\right| \leq \pm\left(\phi (t) -\phi (t_{0})  \right)\right\}\,,
\]
where $\phi (t):= \frac{1}{1-\ell}t^{1-\ell}$ if  $\ell \not=1$, and $\phi (t):= \ln (t+1) $ if $\ell=1$,  is a distance function when $\ell<1$.  
In fact, any intersection of $D_-(x_0, t_0)$ with the hyperplane $0<t = const < t_0$ determines the
so-called {\it dependence domain} for the point $ (x_0, t_0)$, while the intersection of $D_+(x_0, t_0)$
with the hyperplane $t = const > t_0>0$ is the so-called {\it domain of influence} of the point
$ (x_0, t_0)$. 
The Dirac equation (\ref{DE})  is non-invariant with respect to time inversion and its solutions have different properties in different direction of time.
\medskip

In order to distinguish upper and lower 2-spinors we introduce two more $\gamma$-matrices (projection operators),
 the upper-left corner and lower-right corner matrices,
\begin{eqnarray}
\label{gammaUL}
\gamma^U
& = &
\left(
\begin{array}{cc}
 {\mathbb I}_2 & {\mathbb O}_2    \\
 {\mathbb O}_2  & {\mathbb O}_2   \\ 
\end{array}
\right)=\frac{1}{2}({\mathbb I}_4+ \gamma^0 )
,\qquad 
 \gamma^L
  =  \left(
\begin{array}{cc}
 {\mathbb O}_2 & {\mathbb O}_2   \\
{\mathbb O}_2  & {\mathbb I}_2   \\ 
\end{array}
\right)=
\frac{1}{2}({\mathbb I}_4- \gamma^0 )\,.
\end{eqnarray}
Next we define the right co-factor
\begin{eqnarray}
\label{Dcomp}
{\mathscr{D}}^{co} (t,\partial _t,\partial _x)
& := &
i t^{-\frac{\ell }{2}} \gamma ^0 (t^{ i m}  \gamma^U+t^{-i m}\gamma^L )\frac{\partial }{\partial t} 
+i t^{-\frac{3\ell }{2}}  \sum_{k=1}^3\gamma ^k (t^{ i m}  \gamma^U+t^{-i m}\gamma^L )\frac{\partial }{\partial x_k} 
\end{eqnarray}
 of the Dirac operator ${\mathscr{D}} (t,\partial _t,\partial _x) $ of (\ref{DO}), which is chosen such that the composition   
${\mathscr{D}} (t,\partial _t,\partial _x)$ ${\mathscr{D}}^{co} (t,\partial _t,\partial _x) $ is a diagonal matrix of operators.
In fact, the  complementary operator ${\mathscr{D}}^{co} (t,\partial _t,\partial _x) $  
 is not unique. For the Dirac equation in the de~Sitter spacetime such  complementary operators were first suggested in \cite{AP2020}.

Denote  ${\mathcal E}^w(x,t)$ to be the distribution that   
is the fundamental solution to the Cauchy problem for the  wave equation  in the Minkowski spacetime
\begin{equation}
\label{FWE}
{\mathcal E}^w_{ tt} -   \Delta  {\mathcal E}^w  =  0 \,, \quad {\mathcal E}^w(x,0)=\delta (x)\,, \quad {\mathcal E}^w_{t}(x,0)= 0\,.
\end{equation}
Here $\Delta  $ is the Laplace operator in ${\mathbb R}^3$. 
Henceforth, $F \left(  \alpha , \beta ;\gamma ;z\right) $ is the hypergeometric function (see, e.g., \cite{B-E}). Our first main theorem gives for the Dirac operator the fundamental solution (retarded propagator)  with  
support in the  forward cone. 
\begin{theorem} 
\label{TFSDO}
For every    $x_0 \in {\mathbb R}^3$, $ t, t_0 \in {\mathbb R}_+$, the fundamental solution ${\mathcal E}_{+}(x ,t;x_0 ,t_0;m) $ with  
support in the  forward cone  $D_+ (x_0,t_0) $,  that is,  a distribution satisfying 
\begin{eqnarray*}
&  &
\cases{ {\mathscr{D}}(t,\partial _t,\partial _x)
{\mathcal E}_{+}(x ,t;x_0 ,t_0;m)=  \delta ( x-x_0) \delta (t-t_0)  {\mathbb I}_4 ,  \cr
\mbox{\rm supp} \,{\mathcal E}_{+} \subseteq D_+ (x_0,t_0)\,,}
\end{eqnarray*}
is given  as follows
\begin{eqnarray*}
&  &
{\mathcal E}_{+}(x ,t;x_0 ,t_0;m) \\
& = &
- 2  t_0^{\frac{\ell }{2}-i m } {\mathscr{D}}^{co} (t,\partial _t,\partial _x)   \int_0^{ \phi (t)- \phi (t_0 ) }\left (
   \begin{array}{ccccc}     
 E(r,t;t_0; m)    {\mathbb I}_2 & {\mathbb O}_2 \\
   {\mathbb O}_2&   
 E(r,t;t_0; -m)   {\mathbb I}_2   \\ 
   \end{array}
   \right) {\mathcal E}^w(x-x_0,r)\,dr\,,
\end{eqnarray*}
where $\ell \in {\mathbb R}$, $\ell \not=1$, and
\begin{eqnarray}
\label{Edef}
 E(r,t;t_0  ; m) 
& = & 
{ 2^{ \frac{2 im}{ 1-\ell }-1}} (1-\ell )^{\frac{\ell }{1-\ell }}   \phi (t_0  )^{\frac{ \ell+2im}{1-\ell }}
\left( \left(\phi (t)+ \phi (t_0  )  \right)^2- r^2 \right)^{-  \frac{ im}{ 1-\ell }} \\
&  &
\times  F \left(  i \frac{m}{ 1-\ell },  i\frac{m}{ 1-\ell };1;\frac{\left( \phi (t)-  \phi (t_0  )   \right)^2-r^2}{\left( \phi (t)+ \phi (t_0  ) \right)^2-r^2}\right)  \,.  \nonumber 
\end{eqnarray}
\end{theorem}
The fundamental solution to the Cauchy problem is given by the next theorem.

\begin{theorem}
\label{TFSCPDE} 
For every positive $ \varepsilon >0$ and $ t> \varepsilon $ the fundamental solution 
${\mathcal E}_{+}(x ,t;x_0 ;m;\varepsilon )$ to the Cauchy problem, that is, a distribution satisfying 
\begin{eqnarray*}
&  &
\cases{ {\mathscr{D}}(t,\partial _t,\partial _x)
{\mathcal E}_{+}(x ,t;x_0 ;m;\varepsilon )=   {\mathbb O}_4 ,  \cr
{\mathcal E}_{+}(x ,\varepsilon ;x_0 ;m;\varepsilon )=\delta ( x-x_0){\mathbb I}_4\,, }
\end{eqnarray*}
is given as follows
\begin{eqnarray*}
{\mathcal E}_{+}(x ,t;x_0 ;m;\varepsilon )
& = &
-i\varepsilon ^{1+\frac{\ell }{2} -i m} (1-\ell)^{-1}{\mathscr{D}}^{co} (x,t,\partial _t,\partial _x)\gamma^0\\
&  &
\times \int_0^{\phi (t)- \phi (\varepsilon )} \left (
   \begin{array}{ccccc}
K_1 \left(r,t; m ;\varepsilon \right)  {\mathbb I}_2& {\mathbb O}_2   \\
   {\mathbb O}_2& K_1 \left(r,t; -m ;\varepsilon \right) {\mathbb I}_2   \\ 
   \end{array}
   \right){\mathcal E}^w(x-x_0,r) \,dr     ,
\end{eqnarray*} 
where $\ell \in {\mathbb R}$, $\ell \not=1$, and
\begin{eqnarray}
\label{K1def}
&  &
 K_1 \left(r,t; m ;\varepsilon \right)  \\
& := &
 2^{2 i  \frac{m}{1-\ell}} \phi (\varepsilon )^{2 i  \frac{m}{1-\ell}-1}   \left( \left(\phi (t)+ \phi (\varepsilon )  \right)^2-  r ^2\right)^{-  i \frac{m}{1-\ell}}   F \left(  i \frac{m}{1-\ell},  i \frac{m}{1-\ell};1;\frac{\left(\phi (t)- \phi (\varepsilon )\right)^2- r ^2}{\left(\phi (t)+ \phi (\varepsilon )   \right)^2-  r ^2 }\right)  \,. \nonumber 
\end{eqnarray} 
\end{theorem}
In order to write the solution to   the Cauchy problem  
  we introduce the operator  
\begin{eqnarray*}
&  &
{\cal G}(x,t,D_x;m )[f] (x,t)\\
& : =  &
- 2\int_\varepsilon ^{t  }b^{\frac{\ell }{2}-i m } \,d b\int_0^{ \phi (t)- \phi (b ) }  
 E(r,t;b ; m)\int_{{\mathbb R}^3} {\mathcal E}^w(x-y,r)  f( y,b)  \,dy \,dr
, \quad  f \in C_0^\infty({\mathbb R}^{n+1}),
\end{eqnarray*}
and  the operator ${\cal K}_1(x,t,D_x;m;\varepsilon)$   as follows:
\begin{eqnarray*}
&  &
{\cal K}_1(x,t,D_x;m;\varepsilon) [\varphi](x,t) \\
& := &
-i\varepsilon ^{1+\frac{\ell }{2} -i m}(1-\ell)^{-1}
\int_0^{ \phi (t)- \phi (\varepsilon ) }   K_1 \left(r,t; m ;\varepsilon \right)\int_{{\mathbb R}^3} {\mathcal E}^w(x-y,r)  \varphi (y )\,dy\,dr  
\,, \quad \varphi \in C_0^\infty({\mathbb R}^n).
\end{eqnarray*}
\begin{theorem}
\label{CPDE}
The solution to the Cauchy problem
\begin{eqnarray*}
\cases{{\mathscr{D}}(t,\partial _t,\partial _x)\Psi  (x,t)=F( x,t)\,,\quad t>\varepsilon >0\,,\cr
\Psi (x,\varepsilon )=\Psi _\varepsilon (x)\,,}
\end{eqnarray*}
with $m \in {\mathbb C}$, is given as follows
\begin{eqnarray*}
\Psi  (x,t)
& = &
 {\mathscr{D}}^{co} (t,\partial _t,\partial _x)\Bigg\{\left (
   \begin{array}{ccccc}
    {\cal G}(x,t,D_x;m ) {\mathbb I}_2 & {\mathbb O}_2 \\
   {\mathbb O}_2&  {\cal G}(x,t,D_x;-m ){\mathbb I}_2   \\ 
   \end{array}
   \right) [F]  (x,t)\\
   &  &
 + \gamma^0\left (
   \begin{array}{ccccc}
  {\cal K}_1(x,t,D_x;m;\varepsilon ) {\mathbb I}_2& {\mathbb O}_2   \\
   {\mathbb O}_2&  {\cal K}_1(x,t,D_x;-m;\varepsilon ){\mathbb I}_2   \\ 
   \end{array}
   \right) [\Psi  _{\varepsilon }]  (x,t)  \Bigg\},\quad t> \varepsilon >0\,.
\end{eqnarray*} 
\end{theorem}
For the case of $\ell =0 $ see \cite{AP2020}. 

The derivation of the formulas for the fundamental solutions and  for the solution of the Cauchy problem is carried out in three steps. The first step (Section~\ref{compoper}) is  the finding of a complementary operator that reduces the $4\times 4$ system of the first order operators to the  system of the diagonal operator-valued matrix with two pairs of the coinciding non-Fuchsian second-order hyperbolic partial differential operators.  The second step (Section~\ref{SLT}) is  a  reduction of such  second-order hyperbolic partial differential operators to the  generalized  Euler-Poisson-Darboux equation
\begin{eqnarray}
\label{EPDEintq}
&  &
\partial^2_t  u-   A(x,\partial_x)  u+\frac{ 2 i  m }{t }  \partial_t u  =   f\,,
\end{eqnarray}
where  $m \in {\mathbb C}$, $t \in{\mathbb R}_+ $  and $  A(x,\partial_x)$ can be a pseudo-differential operator with the symbol $ A(x,\xi ) $ defined for $(x,\xi) \in \Omega \times {\mathbb R}^n $ and $\Omega  $ is a domain in ${\mathbb R}^n$. The reduction is done  by a change  to the co-moving coordinates.

The last, the third step (Section~\ref{ITA}),  is devoted to  solving  the Cauchy problem for the  generalized  Euler-Poisson-Darboux equation (\ref{EPDEintq}) (Theorems~\ref{T5.4}-\ref{T5.5}). In the  equation  (\ref{EPDEintq})   higher order derivatives with respect to spatial variables may appear. To solve the last equation the integral transform approach from \cite{YagTricomi,JDE2015,Yagdjian-Galstian,MN2015} is employed. In the last two steps it is supposed that $x \in {\mathbb R}^n$ with an arbitrary dimension $n$.

\section{The complementary operator}   
\label{compoper}

The  non-Fuchsian
Type   Dirac operator ${\mathscr{D}} (t,\partial _t,\partial _x)$ in the FLRW spacetime is defined by
\begin{eqnarray*}
{\mathscr{D}}(t,\partial _t,\partial _x)
& := &
i   \gamma ^0 \frac{\partial }{\partial t}   
+i t^{ -\ell } \gamma ^1\frac{\partial }{\partial x_1}
+i t^{ -\ell }  \gamma ^2 \frac{\partial }{\partial x_2} 
+i t^{ -\ell } \gamma ^3\frac{\partial }{\partial x_3}
+i\frac{ 3  }{2 } \ell t^{ -1} \gamma ^0
-m t^{ -1}  {\mathbb I}_4\,.
\end{eqnarray*}
With the aid of  the  matrices $\gamma^U$ and $ \gamma^L $ (\ref{gammaUL}) 
we calculate   
\begin{eqnarray*}
&  &
t^c \gamma^U+\gamma^L=\left (
   \begin{array}{ccccc}
   t^{c}{\mathbb I}_2& {\mathbb O}_2   \\
   {\mathbb O}_2& {\mathbb I}_2   \\ 
   \end{array}
   \right),\quad (t^c  \gamma^U+\gamma^L ) \gamma ^0 (t^r  \gamma^U+\gamma^L )=
\left (
   \begin{array}{ccccc}
   t^{c+r}{\mathbb I}_2& {\mathbb O}_2   \\
   {\mathbb O}_2& -{\mathbb I}_2   \\ 
   \end{array}
   \right)\,,\\
&  &
 (t^c  \gamma^U+\gamma^L ) \gamma ^k (t^r  \gamma^U+\gamma^L )=\left (
   \begin{array}{ccccc}
  {\mathbb O}_2& t^c\sigma ^k   \\
  -t^r\sigma ^k &  {\mathbb O}_2  \\  
   \end{array}
   \right), \qquad k=1,2,3\,.
\end{eqnarray*}
Next we use them to define a family of  complementary factors ${\mathscr{D}}^{co} (t,\partial _t,\partial _x)$.  
For the Dirac operator the auxiliary right complementary factors  to the diagonal  matrix  with the operator-valued entries form a family of operators, which contains  the following operators 
\begin{eqnarray*}
{\mathscr{D}}^{co} (t,\partial _t,\partial _x)
& := &
i t^b (t^c  \gamma^U+\gamma^L )\gamma ^0 \frac{\partial }{\partial t}(t^r  \gamma^U+\gamma^L ) 
+i t^{-\ell } t^b (t^c  \gamma^U+\gamma^L ) \gamma ^1 (t^r  \gamma^U+\gamma^L )\frac{\partial }{\partial x_1}+\\
&  &
+i t^{-\ell } t^b (t^c  \gamma^U+\gamma^L ) \gamma ^2 (t^r  \gamma^U+\gamma^L )\frac{\partial }{\partial x_2}
+i t^{-\ell } t^b (t^c  \gamma^U+\gamma^L ) \gamma ^3(t^r  \gamma^U+\gamma^L )\frac{\partial }{\partial x_3}+\\
&  &
+i\frac{3  }{4 }  b_0 t^{b-1} (t^c  \gamma^U+\gamma^L )\gamma ^0(t^r  \gamma^U+\gamma^L )
+n t^{b-1} (t^c  \gamma^U+\gamma^L )(t^r  \gamma^U+\gamma^L )\,.
\end{eqnarray*}
These operators depend on  parameters $b_0, b,c,n,r \in {\mathbb C}$. We need also a matrix  
\begin{eqnarray*}
t^a  \gamma^U+t^\omega \gamma^L 
& = &
\left(
\begin{array}{cc}
 t^a{\mathbb I}_2 & {\mathbb O}_2    \\
 {\mathbb O}_2  & t^\omega{\mathbb I}_2  \\ 
\end{array}
\right)\,.
\end{eqnarray*}
Then for the parameters $a,b_0,c,n $ chosen  as follows
\begin{equation}
\label{parameters}
a:=\frac{\ell }{2}-i m, \quad \omega {:=}\frac{\ell }{2}+i m,\quad b_0 =-\frac{4 i m}{3},\quad b = -i m-\frac{\ell }{2} ,\quad r =2 i m,\quad c=0, \quad n=m,
\end{equation}
since $\gamma ^U+ \gamma ^L ={\mathbb I}_4$,  we obtain the operator of (\ref{Dcomp})
and the following identity
\begin{eqnarray*}
&  &
{\mathscr{D}}(t,\partial _t,\partial _x) {\mathscr{D}}^{co} (t,\partial _t,\partial _x)\\
& = &
(t^{-a } \gamma ^U+t^{-\omega } \gamma ^L)  \left(
\begin{array}{cc}
- {\mathbb I}_2( \partial_t^2 -  t^{-2\ell } \ell   +\frac{(\ell +2 i m) }{t} \partial_t ) & {\mathbb O}_2   \\
 {\mathbb O}_2 & -{\mathbb I}_2 ( \partial_t^2 -  t^{-2\ell } \ell   +\frac{(\ell -2 i m) }{t} \partial_t )   \\ 
\end{array}
\right)\\
& = &
- (t^{-\frac{\ell }{2}+i m } \gamma ^U+t^{-\frac{\ell }{2}-i m  } \gamma ^L) \Big[ {\mathbb I}_4\partial_t^2 -t^{-2 \ell } {\mathbb I}_4 \ell + t^{-1}\left((\ell+2 i m)  \gamma^U+(\ell-2 i m)\gamma^L \right)\partial_t \Big].
\end{eqnarray*}
If we denote the  second-order scalar hyperbolic operator
\begin{equation}
\label{P}
P(t,\partial_t,\partial_x;m):= \partial_t^2 -  t^{-2\ell } \Delta   +t^{-1}(\ell+2 i m) \partial_t \,,
\end{equation}
then we obtain the identity with the diagonal right-hand side 
\begin{eqnarray}
\label{11}
{\mathscr{D}}(t,\partial _t,\partial _x) {\mathscr{D}}^{co} (t,\partial _t,\partial _x)
& = &
- (t^{-a } \gamma ^U+t^{-\omega } \gamma ^L)\left( \gamma ^UP(t,\partial_t,\partial_x;m)
+ \gamma ^L P(t,\partial_t,\partial_x;-m) \right) \nonumber \\
& = &
- \left(
\begin{array}{cc}
 t^{-a }{\mathbb I}_2P(t,\partial_t,\partial_x;m) & {\mathbb O}_2   \\
 {\mathbb O}_2 & t^{-\omega }{\mathbb I}_2 P(t,\partial_t,\partial_x;-m)   \\ 
\end{array}
\right).
\end{eqnarray}
The  complementary operator ${\mathscr{D}}^{co} (t,\partial _t,\partial _x) $  
 is not unique. An advantage of the choice of parameters $b_0$, $b$, $c$, $n$, $r $  (\ref{parameters}) is that the  diagonal entries of (\ref{11}) are the scalar operators: two  operators $t^{-a }P(t,\partial_t,\partial_x;m) $ and two operators $t^{-\omega }P(t,\partial_t,\partial_x;-m)  $, which  differ only by the sign of the mass  $m$ and the factors $t^{-a } $ and $t^{-\omega } $.
 \medskip

In order to consider the Dirac equation in  curvilinear coordinates as well as to enlarge  class of the partial differential equations,  we follow the   arguments which have been specified in \cite[Sec1]{AP2020}. More exactly, consider  operator 
\begin{equation}
\label{12}
{\mathscr{D}}(x,t,\partial _t,\partial _x)
 := 
i   \gamma ^0 \frac{\partial }{\partial t}   
+i t^{ -\ell } \sum_{k=1}^3\gamma ^k A_{k}(x, \partial_x ) 
+i\frac{ 3  }{2 } \ell t^{ -1} \gamma ^0
-m t^{ -1}  {\mathbb I}_4\,,
\end{equation}
where the operators $A_k(x, \partial_x)  $, $k=1,2,3$,   in general, are the scalar pseudo-differential operators  $A_k(x, \partial_x)$ with symbols depending on the spatial variables too. 
Following  \cite{AP2020}  we    define  the   {\it generalized Dirac operator}  (\ref{12})
as an operator satisfying the  condition 
\begin{equation}
\label{13}
\sum_{k,j=1,2,3}\left(  t^{-i m}\gamma^U + t^{i m}\gamma^L  \right) \gamma ^k   {\gamma }^j   \left( t^{ i m} \gamma^U  + t^{-i m} \gamma^L   \right) A_{k}(x, \partial )A_{j}(x, \partial )=
- {\mathcal A} (x, \partial_x;m){\mathbb I}_4 \quad  \mbox{\rm for all}\,\, t>0\,,
\end{equation}
where ${\mathcal A} (x, \partial_x;m) $ is the scalar pseudo-differential operator independent of $t$.  For $m=0$ this condition coincides with condition (1.17)~\cite{AP2020}. We note that for the pairwise commuting operators this condition is satisfied with 
${\mathcal A} (x, \partial_x;m)= -\sum_{k=1,2,3} A_k^2(x, \partial ) $. 
From now on we   omit $m$ in the notation $ {\mathcal A} (x, \partial_x;m)$ and write just $ {\mathcal A} (x, \partial_x)$. 
\medskip

The  complementary operator ${\mathscr{D}}^{co} (x,t,\partial _t,\partial _x) $ for operator (\ref{12}) is given by the following theorem.  
\begin{theorem}
Assume that for operator (\ref{12}) condition (\ref{13}) is fulfilled.  Then the following operator 
\begin{eqnarray*} 
{\mathscr{D}}^{co} (x,t,\partial _t,\partial _x)
& := &
i t^{-\frac{\ell }{2}} \gamma ^0 (t^{ i m}  \gamma^U+t^{-i m}\gamma^L )\frac{\partial }{\partial t} 
+i t^{-\frac{3\ell }{2}}  \sum_{k=1}^3\gamma ^k (t^{ i m}  \gamma^U+t^{-i m}\gamma^L )A_{k}(x, \partial )
\end{eqnarray*}
 is a complementary operator ${\mathscr{D}}^{co} (x,t,\partial _t,\partial _x) $ for (\ref{12}), that is,
\begin{eqnarray*}
&  &
{\mathscr{D}}(x,t,\partial _t,\partial _x){\mathscr{D}}^{co} (x,t,\partial _t,\partial _x)\\
& = &
- t^{-\frac{\ell }{2}}  (t^{ i m}  \gamma^U+t^{-i m}\gamma^L )\Bigg\{{\mathbb I}_4 \frac{\partial^2 }{\partial t^2}- t^{-2\ell }  
{\mathcal A} (x, \partial_x){\mathbb I}_4 +
 t^{ -1}  (  \ell{\mathbb I}_4 + 2 i m  \gamma ^0 ) \frac{\partial  }{\partial t } 
\Bigg\}\,.
\end{eqnarray*}
\end{theorem}
\medskip

\noindent
{\bf Proof.}  Indeed, it is easy to see that  
\begin{eqnarray*}
{\mathscr{D}}(x,t,\partial _t,\partial _x){\mathscr{D}}^{co} (x,t,\partial _t,\partial _x)
& = & 
I_1+I_2+I_3+I_4\,,
\end{eqnarray*}
where we have denoted
\begin{eqnarray*}
I_1 & = & 
i   \gamma ^0 \frac{\partial }{\partial t} \Bigg\{i t^{-\frac{\ell }{2}} \gamma ^0 (t^{ i m}  \gamma^U+t^{-i m}\gamma^L )\frac{\partial }{\partial t} 
+i t^{-\frac{3\ell }{2}}  \sum_{b=1}^3\gamma ^b (t^{ i m}  \gamma^U+t^{-i m}\gamma^L )A_{b}(x, \partial ) \Bigg\}\,,\\
I_2 & = &   
i t^{ -\ell } \sum_{a=1}^3\gamma ^a A_{a}(x, \partial )   \Bigg\{i t^{-\frac{\ell }{2}} \gamma ^0 (t^{ i m}  \gamma^U+t^{-i m}\gamma^L )\frac{\partial }{\partial t} \\
&  &
+i t^{-\frac{3\ell }{2}}  \sum_{b=1}^3\gamma ^b (t^{ i m}  \gamma^U+t^{-i m}\gamma^L )A_{b}(x, \partial ) \Bigg\}\,,\\
I_3 & =  &  
i\frac{ 3  }{2 } \ell t^{ -1} 
i t^{-\frac{\ell }{2}}   (t^{ i m}  \gamma^U+t^{-i m}\gamma^L )\frac{\partial }{\partial t} 
+i\frac{ 3  }{2 } \ell t^{ -1} i t^{-\frac{3\ell }{2}}  \sum_{b=1}^3\gamma ^0\gamma ^b (t^{ i m}  \gamma^U+t^{-i m}\gamma^L )A_{b}(x, \partial ) \,,\\
I_4 & = & 
-m t^{ -1}  
i t^{-\frac{\ell }{2}} \gamma ^0 (t^{ i m}  \gamma^U+t^{-i m}\gamma^L )\frac{\partial }{\partial t} 
-m t^{ -1}   i t^{-\frac{3\ell }{2}}  \sum_{b=1}^3\gamma ^b (t^{ i m}  \gamma^U+t^{-i m}\gamma^L )A_{b}(x, \partial ) \,.
\end{eqnarray*}
Then it is easy to derive 
\begin{eqnarray*}
I_1 & = & 
-       t^{-\frac{\ell }{2}}  (t^{ i m}  \gamma^U+t^{-i m}\gamma^L )\frac{\partial^2 }{\partial t^2} 
- t^{-\frac{3\ell }{2}}  \sum_{b=1}^3\gamma ^0\gamma ^b (t^{ i m}  \gamma^U+t^{-i m}\gamma^L )A_{b}(x, \partial )   \frac{\partial }{\partial t} \\
&  &
+ \frac{\ell }{2} t^{-\frac{\ell }{2}-1}  (t^{ i m}  \gamma^U+t^{-i m}\gamma^L )\frac{\partial }{\partial t} 
-   i m    t^{-\frac{\ell }{2}-1}  (t^{ i m}  \gamma^U-t^{-i m}\gamma^L )\frac{\partial }{\partial t} \\
&  &
+\frac{3\ell }{2} t^{-\frac{3\ell }{2}-1}  \sum_{b=1}^3\gamma ^0\gamma ^b (t^{ i m}  \gamma^U+t^{-i m}\gamma^L )A_{b}(x, \partial ) \\
&  &
- i mt^{-\frac{3\ell }{2}-1}  \sum_{b=1}^3\gamma ^0\gamma ^b (t^{ i m}  \gamma^U-t^{-i m}\gamma^L )A_{b}(x, \partial )  \,, \\
I_2 
& = &    
-   t^{-\frac{3\ell }{2}}\sum_{a=1}^3 \gamma ^a    \gamma ^0 (t^{ i m}  \gamma^U+t^{-i m}\gamma^L )\frac{\partial }{\partial t}A_{a}(x, \partial )  \\
&  &
-  t^{-\frac{5\ell }{2}}  \sum_{a=1}^3\sum_{b=1}^3\gamma ^a   \gamma ^b (t^{ i m}  \gamma^U+t^{-i m}\gamma^L )A_{a}(x, \partial )A_{b}(x, \partial )\,,\\
I_3 
& =  &   
-\frac{ 3  }{2 } \ell 
 t^{-1-\frac{ \ell }{2}}   (t^{ i m}  \gamma^U+t^{-i m}\gamma^L )\frac{\partial }{\partial t} 
-\frac{ 3  }{2 } \ell   t^{-1-\frac{3\ell }{2}}  \sum_{b=1}^3\gamma ^0\gamma ^b (t^{ i m}  \gamma^U+t^{-i m}\gamma^L )A_{b}(x, \partial )  \,,\\
I_4 
& = & 
-im    t^{-1-\frac{ \ell }{2}} \gamma ^0 (t^{ i m}  \gamma^U+t^{-i m}\gamma^L )\frac{\partial }{\partial t} 
-  im    t^{-1-\frac{3\ell }{2}}  \sum_{b=1}^3\gamma ^b (t^{ i m}  \gamma^U+t^{-i m}\gamma^L )A_{b}(x, \partial )\,.
\end{eqnarray*}
After substitution of all these expressions in ${\mathscr{D}}(x,t,\partial _t,\partial _x){\mathscr{D}}^{co} (x,t,\partial _t,\partial _x)$ 
we obtain operator with the coefficients  as follows. 
The coefficient  of $\partial^2/\partial t^2  $ is 
\[
- t^{-\frac{\ell }{2}}  (t^{ i m}  \gamma^U+t^{-i m}\gamma^L )\,.
\]
The coefficient of the operator $\partial /\partial t    $ is 
\begin{eqnarray*}
&  & \frac{\ell }{2} t^{-\frac{\ell }{2}-1}  (t^{ i m}  \gamma^U+t^{-i m}\gamma^L )
-   i m    t^{-\frac{\ell }{2}-1}  (t^{ i m}  \gamma^U-t^{-i m}\gamma^L ) \\
&  &
-\frac{ 3  }{2 } \ell 
 t^{-1-\frac{ \ell }{2}}   (t^{ i m}  \gamma^U+t^{-i m}\gamma^L )
 -im    t^{-1-\frac{ \ell }{2}} \gamma ^0 (t^{ i m}  \gamma^U+t^{-i m}\gamma^L )\\
 & = &
t^{-\frac{\ell }{2}-1}  (t^{ i m}  \gamma^U+t^{-i m}\gamma^L )\left( -\ell 
-  2 i m \gamma^0  \right)\,.
\end{eqnarray*}
For the terms with  $  A_{a}(x, \partial )\frac{\partial  }{\partial t }$  we obtain 
\begin{eqnarray*}
&  &
- t^{-\frac{3\ell }{2}}  \sum_{b=1}^3\gamma ^0\gamma ^b (t^{ i m}  \gamma^U+t^{-i m}\gamma^L )A_{b}(x, \partial )   \frac{\partial }{\partial t}\\
&  &
-   t^{-\frac{3\ell }{2}}\sum_{a=1}^3 \gamma ^a    \gamma ^0 (t^{ i m}  \gamma^U+t^{-i m}\gamma^L )A_{a}(x, \partial )\frac{\partial }{\partial t}  =0\,.
\end{eqnarray*}
For the terms with $ A_{a}(x, \partial )$ we obtain
\begin{eqnarray*}
\hspace{-0.3cm} &  &
\frac{3\ell }{2} t^{-\frac{3\ell }{2} -1}  \sum_{b=1}^3\gamma ^0\gamma ^b (t^{ i m}  \gamma^U+t^{-i m}\gamma^L )A_{b}(x, \partial ) 
- i mt^{-\frac{3\ell }{2} -1}  \sum_{b=1}^3\gamma ^0\gamma ^b (t^{ i m}  \gamma^U-t^{-i m}\gamma^L )A_{b}(x, \partial )   \\
\hspace{-0.3cm}  &  &
 -\frac{ 3  }{2 } \ell   t^{-\frac{3\ell }{2} -1}  \sum_{b=1}^3\gamma ^0\gamma ^b (t^{ i m}  \gamma^U+t^{-i m}\gamma^L )A_{b}(x, \partial )
 -  im    t^{-\frac{3\ell }{2} -1}  \sum_{b=1}^3\gamma ^b (t^{ i m}  \gamma^U+t^{-i m}\gamma^L )A_{b}(x, \partial )\\
\hspace{-0.3cm}  & &
 =0\,.
\end{eqnarray*}
The remaining terms are 
\begin{eqnarray*}
&  &
-  t^{-\frac{5\ell }{2}}  \sum_{a=1}^3\sum_{b=1}^3\gamma ^a   \gamma ^b (t^{ i m}  \gamma^U+t^{-i m}\gamma^L )A_{a}(x, \partial )A_{b}(x, \partial )\\
& = &
- \left(  t^{i m}\gamma^U + t^{-i m}\gamma^L  \right) \left(  t^{-i m}\gamma^U + t^{i m}\gamma^L  \right)  t^{-\frac{5\ell }{2}}  \sum_{a=1}^3\gamma ^a   \gamma ^b (t^{ i m}  \gamma^U+t^{-i m}\gamma^L )A_{a}(x, \partial )A_{b}(x, \partial )\\
& = &
t^{-\frac{5\ell }{2}} \left(  t^{i m}\gamma^U + t^{-i m}\gamma^L  \right) 
{\mathcal A} (x, \partial_x){\mathbb I}_4\,.
\end{eqnarray*}
Theorem is proved. \qed

\begin{remark}
More general choice of the family of operators has been  suggested in \cite[Sec.1]{AP2020}, which   in the case of $a(t)=a_0t^\ell $ leads to
\begin{eqnarray*} 
&  &
\sum_{k,j=1,2,3}\left(  t^{-i m}\gamma^U + t^{i m}\gamma^L  \right) \gamma ^k   {\gamma }^j   \left( t^{ i m} \gamma^U  + t^{-i m} \gamma^L   \right) A_{k}(x, \partial )A_{j}(x, \partial )\\
& = &
 -   {\mathcal A} (x, \partial_x;m){\mathbb I}_4+{\mathcal B} (x, \partial_x;m){\gamma }^0+{\mathcal C} (x, \partial_x;m)\gamma ^1\gamma ^2+{\mathcal D} (x, \partial_x;m)\gamma ^3\gamma ^0\gamma ^1\gamma ^2\gamma ^3 \quad  \forall \,\, t>0\,,
\end{eqnarray*}
 where $A_k(x, \partial_x)$, $ {\mathcal A} (x, \partial_x;m)$, ${\mathcal B} (x, \partial_x;m) $, ${\mathcal C} (x, \partial_x;m) $,  and ${\mathcal D} (x, \partial_x;m) $ are   scalar pseudo-differential operators.  For the purpose of the present paper it suffices to consider the case of (\ref{13}), that is, ${\mathcal B} (x, \partial_x;m)= $  ${\mathcal C} (x, \partial_x;m) $ $={\mathcal D} (x, \partial_x;m)=0$.  
\end{remark}

 In order to find  a complementary operator for the Dirac operator in the FLRW spacetime in the curvilinear coordinates, we  appeal to the spherical coordinates 
\begin{eqnarray*} 
x(r,\theta ,\phi )
  :=  
r \cos (\phi ) \sin (\theta ),\quad
y(r,\theta ,\phi )
  :=  
r \sin (\phi ) \sin (\theta ),\quad
z(r,\theta ,\phi )
  :=  
r \cos (\theta )
\end{eqnarray*} 
and choose    
a family of operators with the property (\ref{13}) containing the operators with coefficients  
\begin{eqnarray*}
A_1(x, \partial ) 
&  = &
\cos ( \phi ) \sin (\theta)  \partial_ r   
+\frac{ \cos (\phi  ) \cos (\theta)  }{r} \partial_ \theta   -\frac{\sin (\phi ) }{r \sin (\theta )} \partial_ \phi  \, ,
 \\ 
A_2(x, \partial ) 
&  = &
\sin (\phi ) \sin (\theta )  \partial_r  
+\frac{ \sin (\phi ) \cos (\theta )  }{r} \partial_  \theta  
+\frac{\cos (\phi ) }{r \sin (\theta )} \partial_ \phi\,,\\ 
A_3(x, \partial ) 
&  = &  
\cos (\theta )  ( \partial_ r  +V(r))  -\frac{\sin (\theta ) }{r}  \partial_\theta \,,\\
   {\mathcal A} (x, \partial_x) 
& =  &  
 \partial_r^2 +2\frac{1}{r }  (1+r V( r ))\partial_r 
+ \frac{1}{r }  \Big( r V^{\prime}(r )+2V( r ) +r V^2( r  )  \Big)\\
&  &
+\frac{1}{r^2}\Big( \partial_\theta^2 +\cot (\theta ) \partial_\theta+\csc ^2(\theta ) 
\partial_ \phi ^2 \Big)\,.
\end{eqnarray*}
Here $x=(r,\theta,\phi)$ and $V=V(r)$ is a smooth function defined in ${\mathbb R}_+ $.

 We can write the Dirac equation (\ref{DE}) 
as follows
\begin{eqnarray}
\label{DESp}
&  &
 \left(i {\gamma }^0  \partial_0+i   \frac{1}{a(t)}\left( {\gamma }^r_c \partial_r  +{\gamma }_c^\theta  \partial_\theta+ {\gamma }_c^\phi  \partial_\phi  \right)  +i \frac{3\dot a(t)}{2a(t)}    {\gamma }^0     -m t^{-1}{\mathbb I}_4 \right)\Psi (x,t)=F(x,t) \,,
\end{eqnarray}
where in this Cartesian tetrad gauge the gamma matrices will be given by (see, e.g., \cite{Schluter}) 
\begin{eqnarray}
\label{3.14}
{\gamma }^r_c
& = &
\gamma ^1 \cos (\phi ) \sin (\theta )
+\gamma ^2 \sin (\theta ) \sin (\phi ) 
+ \gamma ^3  \cos (\theta )\,,\\
\label{3.15}
{\gamma }^\phi _c 
& = &
 -\gamma ^1\frac{\sin (\phi ) }{r \sin (\theta )} 
+ \gamma ^2 \frac{\cos (\phi ) }{r \sin (\theta )} 
=  \frac{ 1}{r \sin (\theta  )}\left( -\gamma ^1\sin (\phi  ) 
+ \gamma ^2 \cos (\phi  ) \right)  \,,\\
\label{3.16}
{\gamma }^\theta _c 
& = &
\frac{ 1 }{r}\left(\gamma ^1\cos (\theta ) \cos (\phi ) 
+ \gamma ^2\sin (\phi) \cos (\theta  ) 
-\gamma ^3 \sin (\theta  )    \right)\,.
\end{eqnarray}
We have used the subscript $c$ for Cartesian. 
We can also write
\begin{eqnarray*}
&  &
 \left(i {\gamma }^0  \partial_0+i  \frac{1}{a(t)} \left( \tilde{\gamma }_c^r \partial_r 
 +\tilde{\gamma }_c^\phi  \frac{ 1}{r \sin (\theta)} \partial_\phi
+ \tilde{\gamma }_c^\theta  \frac{ 1 }{r} \partial_\theta  \right)  
+i \frac{3\dot a(t)}{2a(t)}    {\gamma }^0     -m t^{-1}{\mathbb I}_4\right)\Psi (x,t)=F(x,t) \,,
\end{eqnarray*} 
where  $\tilde{\gamma }^t_c = {\gamma }^0$ and 
\begin{eqnarray*}
\tilde{\gamma }^r_c
& = &
{\gamma }^r_c=\gamma ^1 \cos (\phi ) \sin (\theta )
+\gamma ^2 \sin (\theta ) \sin (\phi ) 
+ \gamma ^3  \cos (\theta ) \,,\\
\tilde{\gamma }^\phi_c 
& = &
 -\gamma ^1\sin (\phi  ) 
+ \gamma ^2 \cos (\phi  )\,, \\
\tilde{\gamma }^\theta_c 
& = &
\gamma ^1\cos (\theta ) \cos (\phi ) 
+ \gamma ^2\sin (\phi) \cos (\theta  ) 
-\gamma ^3 \sin (\theta  )  \,,    
\end{eqnarray*}
and with the Lorentzian metric $\eta  $ in the Minkowski spacetime  we have
\begin{eqnarray*} 
&  &
\left\{  \tilde{\gamma}^{\mu}_c,  \tilde{\gamma}^{\nu}_c \right\} = 2 \eta ^{\mu \nu} ,\quad \mu ,\nu = t,r,\theta ,\phi,\quad
 \eta ^{r r}=\eta ^{\theta  \theta } =\eta ^{\phi \phi  } =-1,\quad \eta ^{\mu \nu}=0\quad  if \quad \mu \not= \nu\,. 
\end{eqnarray*}
\begin{proposition}
In the spherical coordinates the complementary operator for the Dirac operator in the FLRW spacetime is
\begin{eqnarray*} 
&  &
{\mathscr{D}}^{co} (x,t,\partial _t,\partial _x) \\
& := &
i t^{-\frac{\ell }{2}} \gamma ^0 (t^{ i m}  \gamma^U+t^{-i m}\gamma^L )\frac{\partial }{\partial t}  
+i t^{-\frac{3\ell }{2}}  \left( \tilde{\gamma }_c^r \partial_r 
 +\tilde{\gamma }_c^\phi  \frac{ 1}{r \sin (\theta  )} \partial_\phi 
+ \tilde{\gamma }_c^\theta  \frac{ 1 }{r} \partial_\theta   \right) (t^{ i m}  \gamma^U+t^{-i m}\gamma^L ) \,,
\end{eqnarray*}
that is,
\begin{eqnarray*} 
&  &
{\mathscr{D}}(x,t,\partial _t,\partial _x){\mathscr{D}}^{co} (x,t,\partial _t,\partial _x)\\
&  = &
\Bigg( i   \gamma ^0 \frac{\partial }{\partial t}   
+i t^{ -\ell } \left( \tilde{\gamma }_c^r \partial_r 
 +\tilde{\gamma }_c^\phi  \frac{ 1}{r \sin (\theta  )} \partial_\phi 
+ \tilde{\gamma }_c^\theta  \frac{ 1 }{r} \partial_\theta   \right)
+i\frac{ 3  }{2 } \ell t^{ -1} \gamma ^0
-m t^{ -1}  {\mathbb I}_4 \Bigg)\\
&  &
\times \Bigg(  i t^{-\frac{\ell }{2}} \gamma ^0 (t^{ i m}  \gamma^U+t^{-i m}\gamma^L )\frac{\partial }{\partial t} 
+i t^{-\frac{3\ell }{2}}  \left( \tilde{\gamma }_c^r \partial_r 
 +\tilde{\gamma }_c^\phi  \frac{ 1}{r \sin (\theta  )} \partial_\phi 
+ \tilde{\gamma }_c^\theta  \frac{ 1 }{r} \partial_\theta   \right) (t^{ i m}  \gamma^U+t^{-i m}\gamma^L ) \Bigg)\\
& = &
- t^{-\frac{\ell }{2}}  (t^{ i m}  \gamma^U+t^{-i m}\gamma^L )\Bigg\{-{\mathbb I}_4\frac{\partial^2 }{\partial t^2}
-(\ell{\mathbb I}_4 +2 i m\gamma ^0 )\frac{1}{t}\frac{\partial }{\partial t}\\
&  &
+t^{-2 \ell }\left(   \frac{ \partial^2 }{ \partial r^2}  
+\frac{2}{r } \frac{ \partial }{ \partial r} 
+ \frac{1}{r^2 } \frac{ \partial^2 }{ \partial \theta ^2 }    
+ \frac{\cot (\theta ) }{r^2 } \frac{ \partial }{ \partial \theta  }  
 + \frac{1}{r^2\sin^2 (\theta )} \frac{ \partial^2 }{ \partial \phi ^2  }  \right){\mathbb I}_4\Bigg\}\,. 
\end{eqnarray*}
\end{proposition}
In particular, for the Dirac equation with the source term $ F$ we have 
\begin{eqnarray*}
&  &
i   \partial_t\Psi  +i  \frac{1}{a(t)} \gamma ^0 \left( {\gamma }^r \partial_r + {\gamma }^\phi  \partial_\phi +{\gamma }^\theta  \partial_\theta  \right)\Psi   
+i \frac{3\dot a(t)}{2a(t)}  \Psi     -m t^{-1}\gamma ^0\Psi (x,t)=F(x,t) \,,
\end{eqnarray*}
where in this Cartesian tetrad gauge the gamma matrices are given by (\ref{3.14}),(\ref{3.15}),(\ref{3.16}).

\section{Solution of the Cauchy problem} 

If we want to solve the problem 
\begin{eqnarray}
\label{3}
\cases{{\mathscr{D}}(x,t,\partial _t,\partial _x)\Psi  (x,t)=F( x,t),\quad t>\varepsilon >0\,,\cr
\Psi (x,\varepsilon )=\Psi _\varepsilon (x)\,,}
\end{eqnarray}
where ${\mathscr{D}}(x,t,\partial _t,\partial _x) $ is given by (\ref{12}), then it suffices to solve the scalar  equations given by the operator
\[
P(t,\partial_t,\partial_x;m):= \partial_t^2 -  t^{-2\ell } {\mathcal A} (x, \partial_x)   +t^{-1}(\ell+2 i m) \partial_t \,.
\]
For   ${\mathcal A} (x, \partial_x)= \Delta $ the operator $P(x, t,\partial_t,\partial_x;m)$ coincides with the  second-order scalar hyperbolic operator  (\ref{P}).  Indeed, if we look for the solution in the form
\begin{equation}
\label{WDcoPhi}
\Psi  (x,t)= {\mathscr{D}}^{co} (x,t,\partial _t,\partial _x)\Phi (x,t),
\end{equation}
 then (\ref{3}) implies 
\begin{eqnarray*}
\cases{- \left(
\begin{array}{cc}
 t^{-a }{\mathbb I}_2P(x,t,\partial_t,\partial_x;m) & {\mathbb O}_2   \\
 {\mathbb O}_2 & t^{-\omega }{\mathbb I}_2 P(x,t,\partial_t,\partial_x;-m)   \\ 
\end{array}
\right)\left(
\begin{array}{c}
 \Phi _U( x,t)   \\
\Phi _L( x,t)   \\ 
\end{array}
\right)=\left(
\begin{array}{c}
 F_U( x,t)   \\
 F_L( x,t)   \\ 
\end{array}
\right), \cr
{\mathscr{D}}^{co} (x,\varepsilon ,\partial _t,\partial _x)\Phi  (x,\varepsilon )=\Psi _\varepsilon (x),}
\end{eqnarray*}
where
\[
F( x,t)=\left(
\begin{array}{c}
 F_U( x,t)   \\
 F_L( x,t)   \\ 
\end{array}
\right) , \qquad \Phi  (x,t )=\left(
\begin{array}{c}
 \Phi _U( x,t)   \\
\Phi _L( x,t)   \\ 
\end{array}
\right)\,. 
\]
Hence,
\begin{eqnarray*}
\cases{ 
P(x,t,\partial_t,\partial_x;m) \Phi _U( x,t) =-t^{a } F_U( x,t),\quad t\geq \varepsilon >0,\cr
 P(x,t,\partial_t,\partial_x;-m)\Phi _L( x,t)  = -t^{\omega }F_L( x,t) ,\quad t\geq \varepsilon >0,\cr
{\mathscr{D}}^{co} (x,\varepsilon ,\partial _t,\partial _x)\Phi  (x,t )|_{t=\varepsilon }=\Psi _\varepsilon (x).}
\end{eqnarray*}
Consider the case of $\Psi _\varepsilon (x)=0 $, then  the  functions $ \Phi _U( x,t)$ and $\Phi _L( x,t) $ are  given by Theorem~\ref{T6.1} with  vanishing initial data. For the case of $ F( x,t)=0$ we  choose the zero for the first initial data for the second order equations, that is, $\Phi (x,\varepsilon )=0 $. Then
\begin{eqnarray*}
\left. {\mathscr{D}}^{co} (x,t ,\partial _t,\partial _x)\Phi  (x,t )\right|_{t=\varepsilon}
& = &
\left. i \varepsilon ^{-i m-\frac{\ell }{2}} \gamma ^0 (\varepsilon ^{2 i m}  \gamma^U+\gamma^L )\frac{\partial }{\partial t}  \Phi  (x,t )\right|_{t=\varepsilon}\\
& = &
i \varepsilon ^{-i m-\frac{\ell }{2}}  \left (
   \begin{array}{ccccc}
   \varepsilon ^{2 i m} {\mathbb I}_2& {\mathbb O}_2   \\
   {\mathbb O}_2&  -{\mathbb I}_2     \\ 
   \end{array}
   \right)\left.\frac{\partial }{\partial t}  \left(
\begin{array}{c}
 \Phi _U (x, t)   \\
\Phi _L (x,t )   \\ 
\end{array}
\right) \right|_{t=\varepsilon}  \,.
\end{eqnarray*}
Thus, we have obtained initial conditions for the solutions of the Klein-Gordon equations:
\begin{eqnarray*}
\cases{ 
P(x,t,\partial_t,\partial_x;m) \Phi _U( x,t) =0,\quad t\geq \varepsilon >0,\cr
 P(x,t,\partial_t,\partial_x;-m)\Phi _L( x,t)  = 0 ,\quad t\geq \varepsilon >0,\cr
\dsp \left(
\begin{array}{c}
 \Phi _U (x,\varepsilon )   \\
\Phi _L (x,\varepsilon )   \\ 
\end{array}
\right)  =\left(
\begin{array}{c}
 0   \\
0   \\ 
\end{array}
\right) , \cr
 \left. \dsp \frac{\partial }{\partial t}  \left(
\begin{array}{c}
 \Phi _U (x,t )   \\
\Phi _L (x,t )   \\ 
\end{array}
\right)\right|_{t=\varepsilon} =  \left(
\begin{array}{c}
-i\varepsilon ^{\frac{\ell }{2} -i m}\Psi  _{\varepsilon \,U} (x,\varepsilon )   \\
i\varepsilon ^{\frac{\ell }{2} +i m}\Psi _{\varepsilon \,L} (x,\varepsilon )   \\ 
\end{array}
\right) \,. }
\end{eqnarray*}

The  solution to the last problem  is given by Theorem~\ref{T6.1}.  Let $ A(x,\partial_x)=\sum_{|\alpha | \leq d} a_\alpha (x)\partial_x^\alpha $ be a differential operator with the smooth coefficients $a_\alpha (x) $.
According to Theorem~\ref{T6.1} if   the function $v=v_f  \left(x,r;b  \right) $ solves the problem
\begin{eqnarray*}  
\cases{ 
\dsp  v _{r r } (x,r;b)-     {\mathcal A} (x, \partial_x) v (x,r;b)   =0\,,\cr 
 v(x,0;b )=f(x,b), \quad  v _\tau (x,0;b )= 0\,,} 
\end{eqnarray*}
while  the function $v  =v_\varphi   (x,r ) $ solves the problem
\begin{eqnarray*}  
\cases{ 
\dsp   v _{r r } (x,r )-     {\mathcal A} (x, \partial_x) v (x,r )   =0\,,\cr 
 v(x,0 )=\varphi (x ), \quad  v   (x,0  )= 0\,,} 
\end{eqnarray*}
then the function $u=u(x,t) $  defined by 
\begin{eqnarray*}
  u(x,t )
& = &
 2\int_\varepsilon ^{t  } \,d b\int_0^{ \phi (t)- \phi (b ) }  
 E(r,t;b ; m) v_f  \left(x,r;b  \right) \,dr\\
 &  &
+\int_0^{\phi (t)- \phi (\varepsilon )}    \varepsilon (1-\ell)^{-1} K_1 \left(r,t; m ;\varepsilon \right) v_{\varphi _1}(x,r)\,dr \,,
\end{eqnarray*}
where the kernels $ E(r,t;b ; m)$ and $K_1 \left(r,t; m ;\varepsilon \right) $ are defined by (\ref{Edef}) and (\ref{K1def}), respectively,
is a solution to the Cauchy problem
\begin{eqnarray*}
\cases{ 
  u_{tt} - t^{-2\ell }{\mathcal A} (x, \partial_x)u +t^{-1}(\ell+2 i m)  u_t =   f (x,t)\,,\cr
 u(x,\varepsilon   )=0, \quad  u _t (x,\varepsilon   )= \varphi _1 (x)\,.} 
\end{eqnarray*}
Here  and henceforth in the notations   $v=v_f  \left(x,r;b  \right) $ and  $v  =v_\varphi   (x,r ) $ the subscripts $ f$ and $\varphi $ do  not denote a partial derivative.

Thus, for the 2-spinor solution $ \Phi _U=\Phi _U( x,t)$ of the problem 
\begin{eqnarray*}
\cases{ 
P(x,t,\partial_t,\partial_x;m) \Phi _U( x,t) =-t^{a } F_U( x,t),\quad t\geq \varepsilon >0\,,\cr
\dsp 
 \Phi _U (x,\varepsilon )    =
 0 ,   \quad 
  \dsp (\partial_ t
 \Phi _U) (x,\varepsilon )    =
-i\varepsilon ^{\frac{\ell }{2} -i m}\Psi  _{\varepsilon \,U} (x )  \,,}
\end{eqnarray*}
we obtain
\begin{eqnarray*}
\Phi _U( x,t)
& = &
 - 2\int_\varepsilon ^{t  }b^{\frac{\ell }{2}-i m } \,d b\int_0^{ \phi (t)- \phi (b ) }  
 E(r,t;b ; m)\int_{{\mathbb R}^3} {\mathcal E}^w(x-y,r)  F_U( y,b)  \,dy \,dr\\
 &  &
-i\varepsilon ^{\frac{\ell }{2} -i m}\int_0^{\phi (t)- \phi (\varepsilon )}    \varepsilon (1-\ell)^{-1} K_1 \left(r,t; m ;\varepsilon \right)\int_{{\mathbb R}^3} {\mathcal E}^w(x-y,r)  \Psi  _{\varepsilon \,U} (y )\,dy\,dr ,\quad t> \varepsilon >0\,.
\end{eqnarray*}
Similarly, for the 2-spinor function $ \Phi _L=\Phi _L( x,t)$  we obtain
\begin{eqnarray*}
\Phi _L( x,t)
& = & 
  - 2\int_\varepsilon ^{t  }b^{\frac{\ell }{2}+i m} \,d b\int_0^{ \phi (t)- \phi (b ) }  
 E(r,t;b ; -m)\int_{{\mathbb R}^3} {\mathcal E}^w(x-y,r)  F_L( y,b)  \,dy \,dr\\
 &  &
+i\varepsilon ^{\frac{\ell }{2} +i m}\int_0^{\phi (t)- \phi (\varepsilon )}    \varepsilon (1-\ell)^{-1} 
K_1 \left(r,t; -m ;\varepsilon \right) \int_{{\mathbb R}^3} {\mathcal E}^w(x-y,r)  \Psi  _{\varepsilon \,L} (y )\,dy\,dr ,\quad t> \varepsilon >0\,.
\end{eqnarray*}
Hence, according to (\ref{WDcoPhi}), 
the solution to the Cauchy problem
\begin{eqnarray*}
\cases{{\mathscr{D}}(x,t,\partial _t,\partial _x)\Psi  (x,t)=F( x,t),\quad t>\varepsilon >0\,,\cr
\Psi (x,\varepsilon )=\Psi _\varepsilon (x)\,,}
\end{eqnarray*}
for $  t> \varepsilon >0$ is given as follows
\begin{eqnarray*}
\Psi  (x,t)
& = &
\Bigg\{ i t^{-\frac{\ell }{2}} \gamma ^0 (t^{ i m}  \gamma^U+t^{-i m}\gamma^L )\frac{\partial }{\partial t} 
+i t^{-\frac{3\ell }{2}}  \sum_{l=1}^3\gamma ^l (t^{ i m}  \gamma^U+t^{-i m}\gamma^L )\frac{\partial }{\partial x_l} 
\Bigg\}\\
&  &
\times \Bigg\{ \Bigg[ - 2\int_\varepsilon ^{t  }b^{\frac{\ell }{2}-i m } \,d b\int_0^{ \phi (t)- \phi (b ) }  
 E(r,t;b ; m)\int_{{\mathbb R}^3} {\mathcal E}^w(x-y,r)  F_U( y,b)  \,dy \,dr\\
 &  &
-i\varepsilon ^{\frac{\ell }{2} -i m}\int_0^{\phi (t)- \phi (\varepsilon )}    \varepsilon (1-\ell)^{-1} K_1 \left(r,t; m ;\varepsilon \right)\int_{{\mathbb R}^3} {\mathcal E}^w(x-y,r)  \Psi  _{\varepsilon \,U} (y )\,dy\,dr  \Bigg] \gamma^U \\
&  &
+ \Bigg[  - 2\int_\varepsilon ^{t  }b^{\frac{\ell }{2}+i m} \,d b\int_0^{ \phi (t)- \phi (b ) }  
 E(r,t;b ; -m)\int_{{\mathbb R}^3} {\mathcal E}^w(x-y,r)  F_L( y,b)  \,dy \,dr\\
 &  &
+i\varepsilon ^{\frac{\ell }{2} +i m}\int_0^{\phi (t)- \phi (\varepsilon )}    \varepsilon (1-\ell)^{-1} 
K_1 \left(r,t; -m ;\varepsilon \right) \int_{{\mathbb R}^3} {\mathcal E}^w(x-y,r)  \Psi  _{\varepsilon \,L} (y )\,dy\,dr \Bigg]  \gamma^L\Bigg\}\,.
\end{eqnarray*} 
If we introduce the operator ${\cal G}(x,t,D_x;m ) $ by 
\begin{eqnarray*}
&  &
{\cal G}(x,t,D_x;m )[f](x.t) \\
&  =  &
- 2\int_\varepsilon ^{t  }b^{\frac{\ell }{2}-i m } \,d b\int_0^{ \phi (t)- \phi (b ) }  
 E(r,t;b ; m)\int_{{\mathbb R}^3} {\mathcal E}^w(x-y,r)  f( y,b)  \,dy \,dr
, \quad  f \in C_0^\infty({\mathbb R}^{n+1})\,,
\end{eqnarray*}
and  the operator ${\cal K}_1(x,t,D_x;m;\varepsilon)$   as follows:
\begin{eqnarray*}
&  &
{\cal K}_1(x,t,D_x;m;\varepsilon) [\varphi] (x.t)\\
& = &
-i\varepsilon ^{\frac{\ell }{2} -i m}\int_0^{\phi (t)- \phi (\varepsilon )}    \varepsilon (1-\ell)^{-1} K_1 \left(r,t; m ;\varepsilon \right)  \int_{{\mathbb R}^3} {\mathcal E}^w(x-y,r)  \varphi (y )\,dy\,dr  
\,, \quad \varphi \in C_0^\infty({\mathbb R}^n),
\end{eqnarray*}
then the formula for the solution is 
\begin{eqnarray*}
\Psi  (x,t)
& = &
\Bigg\{ i t^{-\frac{\ell }{2}} \gamma ^0 (t^{ i m}  \gamma^U+t^{-i m}\gamma^L )\frac{\partial }{\partial t} 
+i t^{-\frac{3\ell }{2}}  \sum_{l=1}^3\gamma ^l (t^{ i m}  \gamma^U+t^{-i m}\gamma^L )\frac{\partial }{\partial x_l} 
\Bigg\}\\
&  &
\times \Bigg\{ \Bigg[ {\cal G}(x,t,D_x;m )[F_U]  (x,t)
 + {\cal K}_1(x,t,D_x;m;\varepsilon) [\Psi  _{\varepsilon \,U}]  (x,t)  \Bigg] \gamma^U \\
&  &
+ \Bigg[  {\cal G}(x,t,D_x;-m )[F_U]  (x,t)
-{\cal K}_1(x,t,D_x;-m;\varepsilon) [\Psi  _{\varepsilon \,L}]  (x,t)  \Bigg]  \gamma^L\Bigg\}\\
& = & 
\Bigg\{ i t^{-\frac{\ell }{2}} \gamma ^0 (t^{ i m}  \gamma^U+t^{-i m}\gamma^L )\frac{\partial }{\partial t} 
+i t^{-\frac{3\ell }{2}}  \sum_{l=1}^3\gamma ^l (t^{ i m}  \gamma^U+t^{-i m}\gamma^L )\frac{\partial }{\partial x_l} 
\Bigg\}\\
&  &
\times \Bigg\{\left (
   \begin{array}{ccccc}
    {\cal G}(x,t,D_x;m ) {\mathbb I}_2 & {\mathbb O}_2 \\
   {\mathbb O}_2&  {\cal G}(x,t,D_x;-m ){\mathbb I}_2   \\ 
   \end{array}
   \right) [F]  (x,t)\\
   &  &
 + \gamma^0\left (
   \begin{array}{ccccc}
  {\cal K}_1(x,t,D_x;m;\varepsilon ) {\mathbb I}_2& {\mathbb O}_2   \\
   {\mathbb O}_2&  {\cal K}_1(x,t,D_x;-m;\varepsilon ){\mathbb I}_2   \\ 
   \end{array}
   \right) [\Psi  _{\varepsilon }]  (x,t)  \Bigg\},\quad t> \varepsilon >0\,.
\end{eqnarray*} 
Thus, we have proved   main Theorem~\ref{CPDE}. \qed

\subsection*{Proof of Theorems~\ref{TFSDO}, \ref{TFSCPDE} }

Theorems~\ref{TFSDO},\ref{TFSCPDE} follow  immediately from Theorem~\ref{CPDE}, since the Cauchy problem for the hyperbolic operator ${\mathscr{D}}(t,\partial _t,\partial _x) $ is well posed in the space of distributions defined in   $ {\mathbb R}^3\times (0,\infty)$.

In terms of (\ref{FWE}),   the retarded  fundamental solution (propagator)  ${\mathcal E}_{+}(x ,t;x_0 ,t_0;m) (= {\mathcal E}_{+}(x -x_0,t;0 ,t_0;m))$ with  
support in the  forward cone  $D_+ (x_0,t_0) $, $x_0 \in {\mathbb R}^n$, $t_0 \in {\mathbb R}_+$, \, 
supp$\,{\mathcal E}_{+} \subseteq D_+ (x_0,t_0)$, is given by 
\begin{eqnarray*}
{\mathcal E}_{+}(x ,t;x_0 ,t_0;m) 
& = &
 {\mathscr{D}}^{co} (x,t,\partial _t,\partial _x)\left (
   \begin{array}{ccccc}
    {\cal G}(x,t,D_x;m ) {\mathbb I}_2 & {\mathbb O}_2 \\
   {\mathbb O}_2&  {\cal G}(x,t,D_x;-m ){\mathbb I}_2   \\ 
   \end{array}
   \right) [\delta  _{x_0}\delta _{t_0}]  (x,t)\\
& = &
- 2  t_0^{\frac{\ell }{2}-i m } {\mathscr{D}}^{co} (x,t,\partial _t,\partial _x) \\
&  &
\times  \int_0^{ \phi (t)- \phi (t_0 ) }\left (
   \begin{array}{ccccc}     
 E(r,t;t_0; m)    {\mathbb I}_2 & {\mathbb O}_2 \\
   {\mathbb O}_2&   
 E(r,t;t_0; -m)   {\mathbb I}_2   \\ 
   \end{array}
   \right) {\mathcal E}^w(x-x_0,r)\,dr\,.
\end{eqnarray*}
We set $0<\varepsilon <t_0 $ in the definition of the operator $ {\cal G}(x,t,D_x;m ) $. Hence, for the upper 2-spinor we have
\[
{\cal G}(x,t,D_x;m ) [\delta _{x_0}\delta _{t_0}](x,t) =
- 2  t_0^{\frac{\ell }{2}-i m } \int_0^{ \phi (t)- \phi (t_0 ) }  
 E(r,t;t_0; m) {\mathcal E}^w(x-x_0,r)  \,dr
,\quad t> \varepsilon >0.
\]
Similarly, for the lower 2-spinor we obtain
\[
{\cal G}(x,t,D_x;-m ) [\delta _{x_0}\delta _{t_0}] (x,t) 
 =
- 2  t_0^{\frac{\ell }{2}+i m } \int_0^{ \phi (t)- \phi (t_0 ) }  
 E(r,t;t_0; -m) {\mathcal E}^w(x-x_0,r)  \,dr
,\quad t> \varepsilon >0\,.
\]
Thus, we have proved Theorem~\ref{TFSDO}. \qed

\section{Solution to non-Fuchsian
type hyperbolic partial differential equations}
\label{SLT}

\subsection{Change to the  proper time}

In this and remaining sections we assume that  $x \in {\mathbb R}^n$ with $n \geq 1$. 
Hence, we have to solve  the non-Fuchsian
type hyperbolic partial differential equation 
\begin{eqnarray}
\label{plus}
&  &
  u_{tt} - t^{-2\ell } \Delta u +t^{-1}(\ell+2 i m)  u_t =   f\,,
\end{eqnarray}
where $ \Delta $ is Laplacian in ${\mathbb R}^n$. 
After change to comoving frame  with the inverse function $\phi^{-1} (\varepsilon ) $ if $\ell \not=1 $, 
\[
\phi (t)=\frac{1}{1-\ell  } t^{1-\ell  }, \quad \phi^{-1} (\varepsilon )
= |1-\ell |^{\frac{1}{1-\ell }}\varepsilon ^{\frac{1}{1-\ell }},\qquad \ell \not=1 \,.
\]
We consider the case of $\ell <1$, since the  corresponding modifications for the case of $\ell >1$ are evident. Since we want to solve problem with  the initial data prescribed on  $t=\varepsilon  $, we introduce a new variable $s$,     
\begin{eqnarray*}
&  &
s =\frac{\phi (t)}{ \phi (\varepsilon )}= \left(\frac{t}{\varepsilon }\right)^{1-\ell  }, \quad t=  \left( 1-\ell  \right)^{-\frac{1}{1-\ell }}\varepsilon \phi^{-1} (s)=\varepsilon s ^{\frac{1}{1-\ell }} ,
\end{eqnarray*}
and obtain for the derivatives 
\begin{eqnarray*}
&  & 
\frac{d}{d t} = \frac{ \phi' (t)}{ \phi (\varepsilon )} \frac{d}{d s }=
\frac{1}{ \phi (\varepsilon  )} t^{-\ell}  \frac{d}{d s },\qquad   
\frac{ \phi' (t)}{ \phi (\varepsilon )}=
(1-\ell )^{-\frac{\ell }{1-\ell }}\frac{1}{\phi (\varepsilon )}\left( \phi (t)\right)^{-\frac{\ell }{1-\ell }}\,,\\
&  &
\frac{d^2}{d t^2} = \left( \frac{ \phi' (t)}{ \phi (\varepsilon )} \right)^2 \frac{d^2}{d s^2 }+ \frac{ \phi'{}' (t)}{ \phi (\varepsilon )} \frac{d}{d s }
=\frac{1}{ \phi^2 (\varepsilon )} t^{-2\ell} \frac{d^2}{d s^2 }-\frac{\ell }{ \phi (\varepsilon  )} t^{-\ell-1} \frac{d}{d s }\,.
\end{eqnarray*}
The time $t=\varepsilon  $ corresponds to $s=1$. Then the equation (\ref{plus}) for the function $u=u(x,s)$ reads
\[
u_{ss}(x,s)
- \phi^2 (\varepsilon  ) \Delta u(x,s) + \frac{2 i m }{(1-\ell )s} u_s(x,s) = \varepsilon^{2\ell} \phi^2 (\varepsilon  )s ^{\frac{ 2\ell}{1-\ell }} f(x,\varepsilon s ^{1/(1-\ell  )})\,.
\]
Thus, we arrive at the generalized Fuchsian
type  partial differential equation (or generalized Euler-Poisson-Darboux equation)
\[
\dsp \widetilde u_{\tau \tau }(x,\tau )-    \widetilde A(x,\partial_x) \widetilde u(x,\tau ) +\frac{2 i \widetilde m }{ \tau +1 }\widetilde u_\tau(x,\tau ) = \widetilde f(x,\tau  )
\]
that will be discussed in the next sections. Here  $\widetilde m:= \frac{ m }{1-\ell}$, $\tau=s -1$, and for the case of (\ref{plus}) we have $ A(x,\partial_x)=\Delta $, while in general, 
\begin{eqnarray}
\label{change1} 
&  &
\widetilde A(x,\partial_x)= \phi ^2(\varepsilon )  A(x,\partial_x),\\ 
\label{change2} 
&  &
\widetilde f (x,\tau ) =\left( \frac{ \phi (\varepsilon )}{ \phi' (t)} \right)^2 f(x,\varepsilon (\tau +1) ^{1/(1-\ell  )})\,,\\
\label{change3} 
&  &
\widetilde f (x,\tau )=(1-\ell )^{\frac{2\ell }{1-\ell }}\phi^{\frac{2}{1-\ell }} (\varepsilon ) \left( \tau +1\right)^{\frac{2\ell }{1-\ell }} f(x,\varepsilon (\tau +1) ^{1/(1-\ell  )})\,,\\
\label{change4} 
&  &
t = \varepsilon (\tau +1) ^{1/(1-\ell  )},\quad \tau = \frac{\phi (t)- \phi (\varepsilon )}{ \phi (\varepsilon )}\,,
\end{eqnarray}
where $A(x,\partial_x) $ is a pseudo-differential operator of order $d$. For the Dirac operator written in  Cartesian coordinates  $ A(x,\partial_x)=\Delta $. On the other hand,  the Dirac operator written in other orthogonal coordinate systems is rather complicated partial differential operator with the variable coefficients depending not only on time variable (\ref{DESp}) (see, also, \cite{Schluter,AP2020}). Having this in mind, in what follows, we   discuss more general case (\ref{change1}) of  the  operator $A(x,\partial_x) $  that can be an operator of order $d$ higher than two.

\subsection{Solution in the original time}

To write a solution to the Cauchy problem for the operator $P(t,\partial_t,\partial_t;m )$   we need one more kernel function that is defined by means of $ K_0\left(r,\tau  ; m \right)$ of subsection~\ref{SS5.1}  as follows
\begin{eqnarray}
\label{K0def} 
&  & 
K_0\left(r,t  ; m; \varepsilon \right):=K_0\left( \frac{r}{\phi (\varepsilon )},\frac{\phi (t)- \phi (\varepsilon )}{ \phi (\varepsilon )} ;   \frac{m}{1-\ell}\right)\\
& \!= \!&
-{ 2^{2 i  \frac{m}{1-\ell }}}  \frac{m}{1-\ell }  \phi^{  \frac{2im}{1-\ell }} (\varepsilon )
\left(\left(\phi (t)+ \phi (\varepsilon )\right)^2-r^2\right)^{-  i\frac{m}{1-\ell }}\nonumber \\
&  &
\times \left\{\frac{2 i \left(r^2- \phi  (t)(\phi  (t)- \phi (\varepsilon ))   \right) }{ r^2-\left(\phi (t)- \phi (\varepsilon )\right)^2 } 
F \left(  i \frac{m}{1-\ell },  i \frac{m}{1-\ell };1;\frac{\left( \phi (t)- \phi (\varepsilon ) \right) ^2-r^2}{(  \phi (t)+ \phi (\varepsilon ) )^2-r^2}\right)\right.\nonumber\\
&  &
\left.-\frac{4 i \phi (t)  \phi (\varepsilon )   \left(\phi^2 (t) - \phi^2 (\varepsilon )  -r^2\right)  }{\left(\left(\phi (t)- \phi (\varepsilon )\right)^2- r^2\right) \left((  \phi (t)+ \phi (\varepsilon ) )^2-r^2\right)}\right.\nonumber\\
&  &
\left.\times F \left(  i \frac{m}{1-\ell }+1,  i \frac{m}{1-\ell };1;\frac{\left( \phi (t)- \phi (\varepsilon ) \right) ^2-r^2}{(  \phi (t)+ \phi (\varepsilon ) )^2-r^2}\right)\right\}\,. \nonumber
\end{eqnarray} 

  Since the following results  cover a wide class of  pseudo-differential operators $A(x,\partial_x)$ and even abstract linear operators, we will not specify Sobolev spaces or distributions, and simply assume that these functions possess  continuous partial derivatives in time variable up to second order. 
\begin{theorem}
\label{T6.1}
Assume that the function $v=v_f  \left(x,r;b  \right) \in C_{x,r,b}^{d,2,0}$ solves the problem
\begin{eqnarray*}  
\cases{ 
\dsp  v _{r r } (x,r;b)-     A(x,\partial_x) v (x,r;b)   =0\,,\cr 
 v(x,0;b )=f(x,b), \quad  v _r (x,0;b )= 0\,,} 
\end{eqnarray*}
while  the function $v  =v_\varphi   (x,r )  \in C_{x,r}^{d,2}$ solves the problem
\begin{eqnarray*}  
\cases{ 
\dsp   v _{r r } (x,r )-     A(x,\partial_x) v (x,r )   =0\,,\cr 
 v(x,0 )=\varphi (x ), \quad  v  _r   (x,0  )= 0\,.} 
\end{eqnarray*}
Then the function $u=u(x,t) $  defined by 
\begin{eqnarray*}
&  &
  u(x,t )\\
& = &
 2\int_\varepsilon ^{t  } \,d b\int_0^{ \phi (t)- \phi (b ) }  
 E(r,t;b ; m) v_f  \left(x,r;b  \right) \,dr
+\int_0^{\phi (t)- \phi (\varepsilon )}    \varepsilon (1-\ell)^{-1} K_1 \left(r,t; m ;\varepsilon \right) v_{\varphi _1}(x,r)\,dr\\
&   &
+
\left(\frac{\phi (t)}{ \phi (\varepsilon )} \right)^{-   i\frac{m}{1-\ell} } 
v_{\varphi _0}\left(x,\phi (t)- \phi (\varepsilon ) \right)\\
&  &
+\frac{1}{\phi (\varepsilon )}\int_0^{\phi (t)- \phi (\varepsilon )} \left[ K_0\left(r,t  ; m; \varepsilon \right)+  2 i m   K_1 \left(r,t; m ;\varepsilon \right)  \right]  
v_{\varphi _0}(x, r)\,dr  \,,
\end{eqnarray*}
where   the kernels $ E(r,t;b ; m)$, $ K_1 \left(r,t; m ;\varepsilon \right) $, and $K_0\left(r,t  ; m; \varepsilon \right) $ are defined by (\ref{Edef}), (\ref{K1def}), and (\ref{K0def}), respectively,
is a solution to the Cauchy problem
\begin{eqnarray*}
\cases{ 
  u_{tt} - t^{-2\ell } A(x,\partial_x)u +t^{-1}(\ell+2 i m)  u_t =   f (x,t)\,, \quad t>\varepsilon \,,\cr
 u(x,\varepsilon   )=\varphi _0(x), \quad  u _t (x,\varepsilon   )= \varphi _1 (x)\,.} 
\end{eqnarray*}
\end{theorem}

In order to prove this theorem we consider separately three cases, which correspond to three functions $f(x,t) $, $\varphi _1(x) $, and $\varphi _0 (x)$. 

\subsection{Proof of Theorem~\ref{T6.1} with   $f$. Case of  $\varphi _0=\varphi _1=0 $}

According to Theorem~\ref{T5.4}  
the function
\begin{eqnarray*}
\widetilde u(x,\tau )
& = &
 \int_0^{\tau } \,db\int_0^{\tau -b}   \widetilde  E(r,\tau;b  ;\widetilde  m)  \widetilde v_{\widetilde f}(x,r;b)\,dr \,,
\end{eqnarray*}
where 
\begin{eqnarray*}
\widetilde E(r,\tau  ;b  ;\widetilde m)
& := &
{ 2^{2 i \widetilde m}}(1+b )^{2i\widetilde m}\left( (\tau +b+2 )^2-r^2\right)^{-  i \widetilde m}  F \left(  i \widetilde m,  i \widetilde m;1;\frac{\left(\tau -  b \right)^2-r^2}{\left(\tau +b +2\right)^2-r^2}\right) 
\end{eqnarray*}
and the function $\widetilde v=\widetilde v_{\widetilde f}(x,r;b) \in C_{x,r,b}^{d,2,0}$ is defined by 
\begin{eqnarray*}  
\cases{ 
\dsp \widetilde v_{r r } (x,r;b)-    \widetilde A(x,\partial_x) \widetilde v(x,r;b)   =0\,,\cr 
\widetilde v(x,0;b )=\widetilde f(x,b), \quad \widetilde v_r (x,0;b )= 0\,,} 
\end{eqnarray*}
solves the Cauchy problem
\begin{eqnarray*}  
\cases{ 
\dsp \widetilde u_{\tau \tau }(x,\tau )-    \widetilde A(x,\partial_x) \widetilde u(x,\tau ) +\frac{ 2i \widetilde m}{ \tau +1 }\widetilde u_\tau(x,\tau ) =\widetilde f (x,\tau )\,,\cr 
\widetilde u(x,0 )=0, \quad \widetilde u_\tau (x,0 )= 0\,.} 
\end{eqnarray*}

\begin{lemma}
Assume that the function $z=z_f (x,r;b) \in C_{x,r,b}^{d,2,0}$  solves the problem
\begin{eqnarray*}  
\cases{ 
\dsp  z _{r r } (x,r;b)-     A(x,\partial_x) z (x,r;b)   =0\,,\cr 
 z(x,0;b )=f(x,b), \quad  z _r (x,0;b )= 0\,,} 
\end{eqnarray*}
then 
\begin{eqnarray*}  
\widetilde v_{ \widetilde f}(x,r;b)
& = &
(1-\ell )^{\frac{2\ell }{1-\ell }}\phi^{\frac{2}{1-\ell }} (\varepsilon ) \left( \tau +1\right)^{\frac{2\ell }{1-\ell }} z_f  \left(x,\phi (\varepsilon)r;\varepsilon (b+1)^{1/(1-\ell )} \right) 
\end{eqnarray*}
solves the problem
\begin{eqnarray*}  
\cases{ 
\dsp \widetilde v_{r r } (x,r;b)-    \widetilde A(x,\partial_x) \widetilde v(x,r;b)   =0\,,\cr 
\widetilde v(x,0;b )=\widetilde f(x,b), \quad \widetilde v_r (x,0;b )= 0\,,} 
\end{eqnarray*}
where $\widetilde A(x,\partial_x)= \phi ^2 (\varepsilon ) A(x,\partial_x) $ and
\[
\widetilde f (x,\tau )=(1-\ell )^{\frac{2\ell }{1-\ell }}\phi^{\frac{2}{1-\ell }} (\varepsilon ) \left( \tau +1\right)^{\frac{2\ell }{1-\ell }} f(x,\varepsilon (\tau +1) ^{1/(1-\ell  )})\,.
\]
\end{lemma}
\medskip

\noindent
{\bf Proof.} According to (\ref{change2}) and  (\ref{change3}), 
 we obtain 
\begin{eqnarray*}  
\widetilde v_{\widetilde f}(x,r;b) = (1-\ell )^{\frac{2\ell }{1-\ell }}\phi^{\frac{2}{1-\ell }} (\varepsilon ) \left( b +1\right)^{\frac{2\ell }{1-\ell }} v_{  f}(x,r;b) \,, 
\end{eqnarray*}
where the function $v=v_{ f}(x,r;b) $ is defined by 
\begin{eqnarray*}  
\cases{ 
\dsp  v_{r r } (x,r;b)-    \widetilde A(x,\partial_x) v(x,r;b)   =0\,,\cr 
v(x,0;b )=f(x,\varepsilon (b +1) ^{1/(1-\ell  )}), \quad  v_r (x,0;b )= 0\,.} 
\end{eqnarray*}
Assume that the function $w= w  (x,r;b) \in C_{x,r,b}^{d,2,0}$ solves
\begin{eqnarray*}  
\cases{ 
\dsp  w_{r r } (x,r;b)-    \widetilde A(x,\partial_x) w(x,r;b)   =0\,,\cr 
w(x,0;b )=f(x,b), \quad  w_r (x,0;b )= 0\,,} 
\end{eqnarray*}
then 
\[
 w  (x,r;b) =z (x,\phi (\varepsilon)r;b) \,.
 \]
 Lemma is proved.
\qed
 \bigskip

Consider with $\tau = \frac{\phi (t)- \phi (\varepsilon )}{ \phi (\varepsilon )} $ the function
\begin{eqnarray*}
  u(x,t )
& = &
 \int_0^{\tau } \,db\int_0^{\tau -b}   \widetilde  E(r,\tau;b  ;  \widetilde m)  \widetilde v_{\widetilde f}(x,r;b)\,dr \\
& = &
\int_0^{\frac{\phi (t)- \phi (\varepsilon )}{ \phi (\varepsilon )}  } \,db\int_0^{\frac{\phi (t)- \phi (\varepsilon )}{ \phi (\varepsilon )}  -b}  \widetilde   E \left(r,\frac{\phi (t)- \phi (\varepsilon )}{ \phi (\varepsilon )}  ;b  ; \widetilde  m \right)\\
&  &
\times (1-\ell )^{\frac{2\ell }{1-\ell }}\phi^{\frac{2}{1-\ell }} (\varepsilon ) \left( b +1\right)^{\frac{2\ell }{1-\ell }} z_f  \left(x,\phi (\varepsilon)r;\varepsilon (b+1)^{1/(1-\ell )} \right) \,dr \,.
\end{eqnarray*} 
Using the substitution  $y=\phi (\varepsilon)r $,  we obtain
 \begin{eqnarray*}
  u(x,t )
& = &
\frac{1}{\phi (\varepsilon )}  \int_0^{\frac{\phi (t)- \phi (\varepsilon )}{ \phi (\varepsilon )}  } \,db\int_0^{ \phi (t)- (b+1) \phi (\varepsilon )   }  \widetilde  E \left(\frac{ y}{\phi (\varepsilon )} ,\frac{\phi (t)- \phi (\varepsilon )}{ \phi (\varepsilon )}  ;b  ;\widetilde  m \right)  \\
&  &
\times (1-\ell )^{\frac{2\ell }{1-\ell }}\phi^{\frac{2}{1-\ell }} (\varepsilon ) \left( b +1\right)^{\frac{2\ell }{1-\ell }}
z_f  \left(x,y;\varepsilon (b+1)^{1/(1-\ell  )} \right) \,dy \,.
\end{eqnarray*}
Introducing $\eta =\varepsilon (b+1)^{1/(1-\ell  )} $ and using the following relations
\begin{eqnarray*}
&  &
 b+1= \frac{\phi (\eta )}{ \phi (\varepsilon )},\quad 
  b = \frac{\phi (\eta )-\phi (\varepsilon )}{ \phi (\varepsilon )},\quad db = \frac{\phi' (\eta ) }{ \phi (\varepsilon )}d \eta\,, \\
  &  &
  b=0 \longmapsto \eta =\varepsilon , \quad b=\frac{\phi (t)- \phi (\varepsilon )}{ \phi (\varepsilon )}\longmapsto \eta = t \,,
\end{eqnarray*} 
we obtain
\begin{eqnarray*}
  u(x,t )
& = &
(\phi (\varepsilon ))^{-1} (1-\ell )^{\frac{2\ell }{1-\ell }}\phi^{\frac{2}{1-\ell }} (\varepsilon )  \int_\varepsilon ^{t  } \,d \eta\int_0^{ \phi (t)- \phi (\eta ) } \frac{\phi' (\eta ) }{ \phi (\varepsilon )} \left(\frac{\phi (\eta )}{ \phi (\varepsilon )} \right) ^{2\ell/(1-\ell  )}  \\
&  &
\qquad \times 
\widetilde  E \left( \frac{y}{\phi (\varepsilon )},\frac{\phi (t)- \phi (\varepsilon )}{ \phi (\varepsilon )}  ;\frac{\phi (\eta )-\phi (\varepsilon )}{ \phi (\varepsilon )}  ;\widetilde  m \right)  z_f  \left(x,y;\eta  \right) \,dy.
\end{eqnarray*}
Now we use 
\[
\phi (\eta )=\frac{1}{1-\ell }\eta ^{1-\ell}, \quad  \phi ^\prime(\eta )= \eta ^{-\ell} , \quad  \eta =[(1-\ell )\phi (\eta )]^{\frac{1}{1-\ell }}, \quad  \phi ^\prime(\eta )=  (1-\ell )^{\frac{-\ell}{1-\ell }}\phi (\eta )^{\frac{-\ell}{1-\ell }}\,,
\]
and derive 
\begin{eqnarray*}
  u(x,t )
& = &
(\phi (\varepsilon ))^{-2} (1-\ell )^{\frac{\ell }{1-\ell }}\phi^{\frac{2}{1-\ell }} (\varepsilon )  \int_\varepsilon ^{t  } \,d \eta\int_0^{ \phi (t)- \phi (\eta ) }  \phi (\eta )^{\frac{-\ell}{1-\ell }}   \left(\frac{\phi (\eta )}{ \phi (\varepsilon )} \right) ^{2\ell/(1-\ell  )} \\
&  &
\times 
\widetilde E \left( \frac{y}{\phi (\varepsilon )},\frac{\phi (t)- \phi (\varepsilon )}{ \phi (\varepsilon )}  ;\frac{\phi (\eta )-\phi (\varepsilon )}{ \phi (\varepsilon )}  ;\widetilde  m \right)  z_f  \left(x,y;\eta  \right) \,dy\\
& = &
 (1-\ell )^{\frac{\ell }{1-\ell }}  \int_\varepsilon ^{t  } \phi (\eta )^{\frac{ \ell}{1-\ell }}   \,d \eta\int_0^{ \phi (t)- \phi (\eta ) }  \\
&  &
\times 
\widetilde E \left( \frac{y}{\phi (\varepsilon )},\frac{\phi (t)- \phi (\varepsilon )}{ \phi (\varepsilon )}  ;\frac{\phi (\eta )-\phi (\varepsilon )}{ \phi (\varepsilon )}  ;\widetilde  m \right)  z_f  \left(x,y;\eta  \right) \,dy\,.
\end{eqnarray*}
On the other hand, for the function $\widetilde E$ 
we obtain
\begin{eqnarray*}
&  &
\widetilde E \left( \frac{y}{\phi (\varepsilon )},\frac{\phi (t)- \phi (\varepsilon )}{ \phi (\varepsilon )}  ;\frac{\phi (\eta )-\phi (\varepsilon )}{ \phi (\varepsilon )}  ;\widetilde  m \right) \\
& = &
{ 2^{2 i \widetilde m}}\left( \phi (\eta )  \right)^{2i\widetilde m}\left( \left(\phi (t)+ \phi (\eta )  \right)^2- y^2 \right)^{-  i \widetilde m} F \left(  i \widetilde m,  i \widetilde m;1;\frac{\left( \phi (t)-  \phi (\eta )   \right)^2-y^2}{\left( \phi (t)+ \phi (\eta ) \right)^2-y^2}\right)
\end{eqnarray*}
and, consequently, 
\begin{eqnarray*}
  u(x,t )
& = &
{ 2^{2 i \widetilde m}} (1-\ell )^{\frac{\ell }{1-\ell }}  \int_\varepsilon ^{t  } \phi (\eta )^{\frac{ \ell}{1-\ell }+2i\widetilde m}\,d \eta\int_0^{ \phi (t)- \phi (\eta ) }    
\left( \left(\phi (t)+ \phi (\eta )  \right)^2- y^2 \right)^{-  i \widetilde m}\\
&  &
\times  F \left(  i \widetilde m,  i \widetilde m;1;\frac{\left( \phi (t)-  \phi (\eta )   \right)^2-y^2}{\left( \phi (t)+ \phi (\eta ) \right)^2-y^2}\right) z_f  \left(x,y;\eta  \right) \,dy\,.
\end{eqnarray*}
Hence, according to (\ref{Edef}), the function
\begin{eqnarray*}
  u(x,t )
& = &
2\int_\varepsilon ^{t  } \,d \eta\int_0^{ \phi (t)- \phi (\eta ) }  
 E(y,t;\eta ; m) z_f  \left(x,y;\eta  \right) \,dy
\end{eqnarray*}
solves equation (\ref{plus}) in accordance to the integral transform approach of \cite{YagTricomi,JDE2015}. Thus, we have proved the case of  $\varphi _0=\varphi _1=0 $ of Theorem~\ref{T6.1}.

\subsection{Proof of Theorem~\ref{T6.1} with $ \varphi _1$. Case of   $f=0$, $\varphi _0=0 $ }

According to Theorem~\ref{T5.4}, the function
\begin{eqnarray*}
\widetilde u(x,\tau )
& = &
\int_0^{\tau }   K_1(r,\tau  ; \widetilde  m)  \widetilde  v_{\varphi _1}(x,r)\,dr
\end{eqnarray*}
solves the Cauchy problem
\begin{eqnarray*}  
\cases{ 
\dsp \widetilde u_{\tau \tau }(x,\tau )-    \widetilde A(x,\partial_x) \widetilde u(x,\tau ) +\frac{ 2 i \widetilde  m}{ \tau +1 }\widetilde u_\tau(x,\tau ) =0\,,\cr 
\widetilde u(x,0 )=0, \quad \widetilde u_\tau (x,0 )= \varphi _1(x)\,.} 
\end{eqnarray*}
where $ \widetilde v=\widetilde v_{\varphi}(x,r)$ is a solution to the problem
\begin{eqnarray*} 
 \cases{
  \widetilde v_{rr}(x,r ) -    \widetilde A(x,\partial_x)\widetilde v(x,r )    =0 \,,\cr 
\widetilde v (x,0 )= \varphi (x), \quad  \widetilde v_r (x,0 )= 0 \,.}
\end{eqnarray*}
Note that $ v_\varphi (x,r)=\widetilde v_{\varphi}(x,\phi^{-1}  (\varepsilon )r)$ solves the equation  $  v_{rr}(x,r ) -      A(x,\partial_x)v(x,r )    =0  $. 
We set $\tau = \frac{\phi (t)- \phi (\varepsilon )}{ \phi (\varepsilon )}$ (\ref{change4})  and consider  the function
\begin{eqnarray*}
  u(x,t )
& = &
 \frac{\varepsilon}{1-\ell} \int_0^{\tau }   K_1(r,\tau  ; \widetilde  m)   \widetilde v_{\varphi _1}(x,r)\,dr \\
& = &
 \frac{\varepsilon}{1-\ell} \int_0^{\frac{\phi (t)- \phi (\varepsilon )}{ \phi (\varepsilon )}}   K_1 \left(r,\frac{\phi (t)- \phi (\varepsilon )}{ \phi (\varepsilon )} ; \widetilde  m \right)   \widetilde v_{\varphi _1}\left(x, r\right)\,
d r\\
& = &
 \frac{\varepsilon}{1-\ell} \int_0^{\frac{\phi (t)- \phi (\varepsilon )}{ \phi (\varepsilon )}}   K_1 \left(r,\frac{\phi (t)- \phi (\varepsilon )}{ \phi (\varepsilon )} ; \widetilde  m \right) v_{\varphi _1} (x,\phi (\varepsilon )r) \,
d r\\
& = &
 \frac{\varepsilon}{1-\ell} \frac{1 }{ \phi (\varepsilon )} \int_0^{\phi (t)- \phi (\varepsilon )}   K_1 \left(\frac{1 }{ \phi (\varepsilon )} y,\frac{\phi (t)- \phi (\varepsilon )}{ \phi (\varepsilon )} ; \widetilde  m \right)  v_{\varphi _1} (x,y) \,
d y\\
& = &
 \frac{\varepsilon}{1-\ell} \frac{1 }{ \phi (\varepsilon )} \int_0^{\phi (t)- \phi (\varepsilon )}   2^{2 i  \widetilde  m}  \phi (\varepsilon )^{2 i  \widetilde  m}\left( \left(\phi (t)+ \phi (\varepsilon  ) \right)^2-  y ^2\right)^{-  i\widetilde   m}  \\
 &  &
 \times F \left(  i \widetilde  m,  i \widetilde  m;1;\frac{\left(\phi (t)- \phi (\varepsilon )\right)^2- y ^2}{\left(\phi (t)+ \phi (\varepsilon   )\right)^2-  y ^2 }\right)  v_{\varphi _1} (x,y) \,
d y\,, \quad t\geq \varepsilon>0\,. 
\end{eqnarray*}
  Finally,
\begin{eqnarray*}
  u(x,t )
& = &
 2^{2 i  \widetilde  m} \phi (\varepsilon )^{2 i  \widetilde  m-1} \frac{\varepsilon}{1-\ell}  \int_0^{\phi (t)- \phi (\varepsilon )}   \left( \left(\phi (t)+ \phi (\varepsilon )  \right)^2-  y ^2\right)^{-  i \widetilde  m}  \\
 &  &
 \times F \left(  i \widetilde  m,  i \widetilde  m;1;\frac{\left(\phi (t)- \phi (\varepsilon )\right)^2- y ^2}{\left(\phi (t)+ \phi (\varepsilon)   \right)^2-  y ^2 }\right)  v_{\varphi _1} (x,y) \,
d y\,,\\
& = &
 \int_0^{\phi (t)- \phi (\varepsilon )} \frac{\varepsilon}{1-\ell}   K_1 \left(r,t; m ;\varepsilon \right) v_{\varphi _1} (x,r) \,
d r\,,  \quad t\geq \varepsilon>0 \,,
\end{eqnarray*}
where  
(\ref{K1def})  has been used.
Thus, we have proved the case of  $\varphi _0 (x)=0 $ and $f(x,t)=0$ of Theorem~\ref{T6.1}.

\subsection{Proof of Theorem~\ref{T6.1} with $ \varphi _0$. Case of  $f=0$, $\varphi _1=0 $ }

According to Theorem~\ref{T5.4},  
if the function $v= v_{\varphi _0}(x,r )\in C_{x,r}^{d,2}$ solves the Cauchy problem
\begin{eqnarray*}  
\cases{  
 \widetilde  v_{rr}(x,r )-  \widetilde A(x,\partial_x)\widetilde  v (x,r )  =0 \,,\cr 
\widetilde v(x,0 )=\varphi _0(x), \quad \partial_r \widetilde v (x,0 )= 0\,, }
\end{eqnarray*}
then the function
\[
\widetilde u(x,\tau )
  =  
(1+ \tau )^{- 3 i  \widetilde m } \widetilde v_{\varphi _0}(x,\tau )+ \int_0^{\tau } \Big[ K_0(r,\tau  ;  \widetilde m)+  2 i   \widetilde m  K_1(r,\tau  ; \widetilde m)\Big] \widetilde v_{\varphi _0}(x,r)\,dr
\]
solves the Cauchy problem
\begin{eqnarray*}  
\cases{ 
\dsp \widetilde u_{\tau \tau }(x,\tau )-    \widetilde A(x,\partial_x) \widetilde u(x,\tau ) +\frac{ 2 i   \widetilde m}{ \tau +1 }\widetilde u_\tau(x,\tau ) =0 \,,\cr
\widetilde u(x,0 )=\varphi _0(x), \quad \widetilde u_\tau (x,0 )= 0\,.} 
\end{eqnarray*}
Note that $ v_\varphi (x,r)=\widetilde v_{\varphi}(x,\phi^{-1}  (\varepsilon )r)$ solves the equation $  v_{rr}(x,r ) -      A(x,\partial_x)v(x,r )    =0 $. 
Set $\tau =\frac{\phi (t)- \phi (\varepsilon )}{ \phi (\varepsilon )} $  (\ref{change4}), then the function
\begin{eqnarray*} 
&  &
 u (x,t )\\
& = &
(1+ \tau )^{-   i   \widetilde m } \widetilde v_{\varphi _0}(x,\tau )+ \int_0^{\tau } \Big[ K_0(r,\tau  ;   \widetilde m)+  2 i   \widetilde m  K_1(r,\tau  ;   \widetilde m)\Big]\widetilde  v_{\varphi _0}(x,r)\,dr  \\
& = &
\left(\frac{\phi (t)}{ \phi (\varepsilon )} \right)^{-   i   \widetilde m } 
\widetilde v_{\varphi _0}\left(x,\frac{\phi (t)- \phi (\varepsilon )}{ \phi (\varepsilon )} \right)\\
&  &
+\int_0^{\frac{\phi (t)- \phi (\varepsilon )}{ \phi (\varepsilon )}} \left[ K_0\left(r,\frac{\phi (t)- \phi (\varepsilon )}{ \phi (\varepsilon )} ;   \widetilde m\right)+  2 i   \widetilde m  K_1\left(r,\frac{\phi (t)- \phi (\varepsilon )}{ \phi (\varepsilon )}  ;   \widetilde m \right)  \right] \widetilde v_{\varphi _0}(x,r)\,dr  \\
& = &
\left(\frac{\phi (t)}{ \phi (\varepsilon )} \right)^{-   i   \widetilde m } 
v_{\varphi _0}\left(x,\phi (t)- \phi (\varepsilon ) \right)\\
&  &
+\int_0^{\frac{\phi (t)- \phi (\varepsilon )}{ \phi (\varepsilon )}} \left[ K_0\left(r,\frac{\phi (t)- \phi (\varepsilon )}{ \phi (\varepsilon )} ;   \widetilde m\right)+  2 i   \widetilde m  K_1\left(r,\frac{\phi (t)- \phi (\varepsilon )}{ \phi (\varepsilon )}  ;   \widetilde m \right)  \right]  v_{\varphi _0}(x, \phi (\varepsilon )r)\,dr \\
& = &
\left(\frac{\phi (t)}{ \phi (\varepsilon )} \right)^{-   i   \frac{m}{1-\ell} } 
v_{\varphi _0}\left(x,\phi (t)- \phi (\varepsilon ) \right)
+\frac{1}{\phi (\varepsilon )}\int_0^{\phi (t)- \phi (\varepsilon )} \Bigg[ K_0\left( \frac{y}{\phi (\varepsilon )},\frac{\phi (t)- \phi (\varepsilon )}{ \phi (\varepsilon )} ;   \frac{m}{1-\ell}\right) \\
&  &
+    \frac{2 im}{1-\ell} K_1\left( \frac{y}{\phi (\varepsilon )},\frac{\phi (t)- \phi (\varepsilon )}{ \phi (\varepsilon )}  ;   \frac{m}{1-\ell} \right)  \Bigg]  
v_{\varphi _0}(x, y)\,dy 
\end{eqnarray*}
solves the problem
\begin{eqnarray*}
\cases{ 
  u_{tt} - t^{-2\ell } A(x,\partial_x)u +t^{-1}(\ell+2 i m)  u_t =  0\,,\cr
 u(x,\varepsilon   )=\varphi _0(x), \quad  u _t (x,\varepsilon   )= 0\,.} 
\end{eqnarray*}
Thus, we have proved the case of  $\varphi _1 (x)=0 $ and $f(x,t)=0$ of Theorem~\ref{T6.1}.

\section{Integral transform approach to generalized  \\ Euler-Poisson-Darboux equation}
\label{ITA}

From now on in this section we will omit ``tilde'' in the notations. The rest of this section is devoted to the following  generalized Euler-Poisson-Darboux equation  
\begin{eqnarray}
\label{EPDEq}
&  &
\partial^2_\tau  u-   A(x,\partial_x)  u+\frac{ 2 i  m }{\tau+1 }  \partial_ \tau u  =   f\,,
\end{eqnarray}
where  $m \in {\mathbb C}$  and $  A(x,\partial_x)$ is a pseudo-differential operator with the symbol $ A(x,\xi ) $ defined for $(x,\xi) \in \Omega \times {\mathbb R}^n $. Here $\Omega  $ is a domain in ${\mathbb R}^n$. For the sake of the previous sections it is enough to set $\Omega ={\mathbb R}^n $. For the  cases with $\Omega \not=  {\mathbb R}^n$ one can consult \cite{JDE2015}.

There is a very extensive literature on  Euler-Poisson-Darboux equation, that is equation (\ref{EPDEq}) with $A(x,\partial_x)$ $=\Delta $ (see, e.g., \cite{Bitsadze,Carroll-Showalter,Delache-Leray, Diaz-Weinberger,Parenti-Tahara,Shishkina,Weinstein,Wirth,JDE2015} and bibliography therein). In particular,    Wirth~\cite{Wirth} used Bessel functions to represent the solution of Euler-Poisson-Darboux equation. In \cite{Palmieri} the version of  the  integral transform approach  \cite{YagTricomi,Yagdjian-Galstian} was employed by  Palmieri   for the derivation of the integral representation formulas for the solution of a linear  equation with the  coefficients, which are scale-invariant and independent  of $x$ variable.  Then Palmieri and coauthors \cite{HHPalmieri} applied these formulas to examine the  blowup phenomena for the wave equations with different types of nonlinearities in the spacetime with power type  expansion.

By the integral transform approach suggested in  \cite{YagTricomi,JDE2015,Yagdjian-Galstian,MN2015} several results presented in the literature can be extended to the  generalized Euler-Poisson-Darboux equation (\ref{EPDEq}). Here we present one of such generalizations.

\subsection{Kernels for Generalized  Euler-Poisson-Darboux equation.}
\label{SS5.1}
To solve  the Cauchy problem for (\ref{EPDEq})   
with $m \in {\mathbb C}$ and data on $\tau  =0$ we use  in this section the following kernel functions
\begin{eqnarray*}
E(r,\tau  ;b  ;m)
& \!= \!&
{ 2^{2 i  m}}(1+b)^{2im}\left( (\tau +b+2 )^2-r^2\right)^{-  i m}  F \left(  i m,  i m;1;\frac{\left(\tau -  b \right)^2-r^2}{\left(\tau +b +2\right)^2-r^2}\right)\,,\\
K_1(r,\tau  ; m)
& \!= \!&
 E(r,\tau   ;0  ;m)={ 2^{2 i  m}}\left( (\tau +2 )^2-r^2\right)^{-  i m}  F \left(  i m,  i m;1;\frac{\tau ^2-r^2}{\left(\tau  +2\right)^2-r^2}\right)\,,\\
K_0(r,\tau  ; m)
& \!= \!&
\lim_{b  \to 0} \left(- \frac{\partial }{\partial b } E(r,\tau  ;b  ;m)\right)\\
&\! = \!&
-{ 2^{2 i  m}}  m \left((\tau +2)^2-r^2\right)^{-  i m}\\
&  &
\times  \left(\frac{2 i \left(r^2-\tau  (\tau +1)\right) }{ r^2-\tau^2 } F \left(  i m,  i m;1;\frac{\tau ^2-r^2}{(\tau +2)^2-r^2}\right)\right.\\
&  &
\left.-\frac{4 i (\tau +1) \left(\tau  (\tau +2)-r^2\right)  }{\left(\tau ^2-r^2\right) \left((\tau +2)^2-r^2\right)}F \left(  i m+1,  i m;1;\frac{\tau ^2-r^2}{(\tau +2)^2-r^2}\right)\right)\,.
\end{eqnarray*} 
The statements of the next theorem are the consequences of  Theorem~2.12~\cite{MN2015}, Propositions~2.14, 2.15 \cite{MN2015}.
\begin{theorem}
\label{MNT2.12}
The functions $E $, $K_0  $, and $K_1 $, solve  the Euler-Poisson-Darboux equation, that is,
\begin{eqnarray}
\label{Eeq}
&  & 
E_{\tau \tau }(r,\tau  ;b  ;m)-    E_{rr}(r,\tau  ;b  ;m)+\frac{ 2 i m}{ \tau +1 }E_\tau(r,\tau  ;b  ;m) =0\,,\\
\label{K0eq}
&  & 
K _{0 \, \tau \tau }(r,\tau  ; m)-    K_{0 \,rr}(r,\tau  ; m)+\frac{ 2 i m}{ \tau +1 }K_{0 \,\tau}(r,\tau  ; m) =0\,,\\
\label{K1eq}
&  & 
K _{1 \, \tau \tau }(r,\tau  ; m)-    K_{1 \,rr}(r,\tau  ; m)+\frac{ 2 i m}{ \tau +1 }K_{1 \,\tau}(r,\tau  ; m) =0\,,
\end{eqnarray}
respectively.
\end{theorem}
\medskip

\noindent
{\bf Proof.} We prove the statement for the function $E$ only, since for two remaining functions the statements follow from the first  one. According to Theorem~2.12~\cite{MN2015},  
the function  
\begin{eqnarray*}
W(r,\tau ;b;M) 
& :=  &
 4 ^{-M}  (b\tau  )^{ -M} \Big((b+\tau )^2 - r^2\Big)^{M -\frac{1}{2}}
 F\Big(\frac{1}{2}-M   ,\frac{1}{2}-M  ;1;
\frac{ ( b-\tau )^2 -r^2 }{( b+\tau )^2 -r^2 } \Big) 
\end{eqnarray*}
solves the following  linear partial differential equation with parameters $b$ and $M$: 
 \begin{eqnarray*}
&  &
W_{\tau \tau } -   W_{rr} +\frac{1}{\tau }W_{\tau  }  -\frac{1}{\tau^2 }M^2 W  =0\,.
\end{eqnarray*}
Furthermore,   for the function $W=\tau ^{-M} V $ we obtain
\[
V_{\tau \tau }  -    V_{rr} +\frac{1-2M }{\tau }V_{\tau  } 
  =0\,.
\]
Hence with $1-2M=2im$ 
\[
\tau ^{- im+\frac{1}{2}}W\left(r,\tau ;b; im-\frac{1}{2} \right) = V \left(r,\tau ;b; im-\frac{1}{2} \right)
\]
and the function
\[
V \left(r,\tau ;b; im-\frac{1}{2} \right)
  =  
2 ^{2im-1}  (b )^{ -(\frac{1}{2}-  im)} \Big((b+\tau )^2 - r^2\Big)^{-  im} 
F\Big(  im   , im  ;1;
\frac{ ( b-\tau )^2 -r^2 }{( b+\tau )^2 -r^2 } \Big)
\]
solves the equation
\[
V_{\tau \tau }  -    V_{rr} +\frac{2im }{\tau }V_{\tau  } 
  =0\,.
\]
It remains to shift $\tau \longrightarrow \tau +1 $ and $ b \longrightarrow b+1$. 
Consequently, the function
\[
E(r,\tau  ;b  ;m)
  =  
{ 2^{2 i \widetilde m}}(1+b )^{2im}\left( (\tau +b+2 )^2-r^2\right)^{-  i m}  F \left(  i m,  i m;1;\frac{\left(\tau -  b \right)^2-r^2}{\left(\tau +b +2\right)^2-r^2}\right)
\]
solves equation (\ref{Eeq}). Theorem is proved.
\qed

\begin{lemma}
\label{L7.3}
We have 
\begin{eqnarray*}
&  &
\lim_{\tau \to 0} 
   K_1(\tau ,\tau  ; m)   = 1\,, \\
&  &
\lim_{\tau \to 0} \left( K_0(\tau ,\tau  ; m)+  2 i m  K_1(\tau ,\tau  ; m) \right)=  i  m\,.
\end{eqnarray*}
\end{lemma}
The proof of the lemma is omitted. The next proposition is an analog of Propositions~2.9,~2.13,~2.15~\cite{MN2015} and we skip its proof.
\begin{proposition} 
\label{MN_P2.9} 
The following hold for the kernel functions
\begin{eqnarray*}
&  &
 E(0,\tau;\tau   ; m)
  =  
 1\,,\\
&  &
 E_r(\tau -b,\tau;b  ; m)    
+   
       E_\tau(\tau -b,\tau;b  ; m)  
+\frac{   i m}{ \tau +1 }      E(\tau -b,\tau;b  ; m) =0\,,\\
&   &
   K_{1\,r}(\tau ,\tau  ; m)   +   K_{1\,\tau }(\tau ,\tau  ; m)    
+  \frac{   i m}{ \tau +1 }  K_1(\tau ,\tau  ; m) =0\,,\\
& &
2 K_{0\,r} (\tau ,\tau  ; m)  + 2K_{0\,\tau } (\tau ,\tau  ; m) +  \frac{ 2 i m}{ \tau +1 }   K_0(\tau ,\tau  ; m)=
-    m (  m+i) (\tau +1)^{-2-  i m}\,.
\end{eqnarray*}
\end{proposition}

\subsection{Solution with $f $, $\varphi _1 $, $\varphi _0 $ for equation in the proper  time}

\begin{theorem}
\label{T5.4}
Let the  function $v= v_{f}(x,r;b ) \in C_{x,r,b}^{d,2,0}$ be a solution to the Cauchy problem
\begin{eqnarray*}  \cases{
  v_{rr}(x,r;b ) -    A(x,\partial_x)v(x,r;b )    =0 \,,\cr 
v (x,0;b )=  f(x,b), \quad  v_r (x,0;b )= 0 \,,}
\end{eqnarray*}
while $ v=v_{\varphi}(x,r) \in C_{x,r}^{d,2}$ is a solution to the problem
\begin{eqnarray*} 
 \cases{
  v_{rr}(x,r ) -    A(x,\partial_x)v(x,r )    =0 \,,\cr 
v (x,0 )= \varphi (x), \quad  v_r (x,0 )= 0 \,.}
\end{eqnarray*}
Then the function
\begin{eqnarray*}
u(x,\tau )
& = &
\int_0^{\tau } \,db\int_0^{\tau -b}  E(r,\tau;b  ;  m)   v_{f}(x,r;b)\,dr +  \int_0^{\tau }   K_1(r,\tau  ;  m)   v_{\varphi _1}(x,r)\,dr\\
&  &
+ (1+ \tau )^{-   i  m } v_{\varphi _0}(x,\tau )+ \int_0^{\tau } \Big[ K_0(r,\tau  ;  m)+  2 i  m  K_1(r,\tau  ;  m)\Big] v_{\varphi _0}(x,r)\,dr
\end{eqnarray*}
solves the Cauchy problem
\begin{eqnarray*}  
\cases{ 
\dsp \partial_\tau^2  u(x,\tau )-     A(x,\partial_x) u(x,\tau ) +\frac{ 2 i  m}{ \tau +1 } u_\tau(x,\tau ) = f(\tau ,x)\,, \cr
 u(x,0 )=\varphi _0(x), \quad  u_\tau (x,0 )= \varphi _1(x)\,.} 
\end{eqnarray*}
\end{theorem}
We stress here that the operator $A(x,\partial_x) $ is a pseudo-differential operator  without any restriction on its order $d$ or type. Thus, there is no any assumption on the type of equation (\ref{EPDEq}). Furthermore, the source term and initial data may be chosen from the Sobolev spaces as well. 

If we assume that $  A(x,\partial_x)$ is an elliptic operator of the second order, then equation (\ref{EPDEq}) is strictly hyperbolic. Therefore,    from the last theorem and from the well-posedness of the Cauchy problem for the hyperbolic operator in the domain $t>0$ follow the next statements about the fundamental solutions. 

\begin{theorem}
Let  ${\mathcal E}_{A}$ be a fundamental solution  of the Cauchy problem
\begin{eqnarray*} 
 \cases{
\partial_r^2  {\mathcal E}_{ A}(x,r;x_0 ) -    A(x,\partial_x) {\mathcal E}_{ A}(x,r;x_0 )    =0 \,,\cr 
 {\mathcal E}_{ A}(x,0 ;x_0  )= \delta  (x-x_0), \quad \partial_r  {\mathcal E}_{ A} (x,0;x_0  )= 0 \,,}
\end{eqnarray*}
then the distribution with the support in $\left\{(x,\tau )\,|\, x \in {\mathbb R}^n, \tau \geq \tau _0\right\}$ that is defined by
\[
{\mathcal E}_{+}(x,\tau ;x_0,\tau_0 )
=
\int_0^{\tau -\tau_0}  E(r,\tau;\tau_0  ;  m)  {\mathcal E}_{ A}(x,r;x_0 )  \,dr \quad \mbox{if} \quad \tau  >\tau_0>0\,,
\]
solves the equation
\[
\dsp \partial_\tau^2 {\mathcal E}_{+}(x,\tau ;x_0,\tau_0 )-  A(x,\partial_x) {\mathcal E}_{+}(x,\tau ;x_0,\tau_0 ) +\frac{ 2 i m}{ \tau +1 }\partial_\tau{\mathcal E}_{+}(x,\tau ;x_0,\tau_0 )= \delta (x-x_0)\delta (\tau -\tau _0). 
\]
\end{theorem}

\begin{theorem}
\label{T5.5}
Let  ${\mathcal E}_{A}$ be a fundamental solution  of the Cauchy problem, that is
\begin{eqnarray*} 
 \cases{
\partial_r^2  {\mathcal E}_{A}(x,r;x_0 ) -   A(x,\partial_x) {\mathcal E}_{A}(x,r;x_0 )    =0 \,,\cr 
 {\mathcal E}_{A}(x,0 ;x_0  )= \delta  (x-x_0), \quad \partial_r  {\mathcal E}_{A} (x,0;x_0  )= 0 \,,}
\end{eqnarray*}
then the distributions
\begin{eqnarray*}
{\mathcal E}_{0}(x,\tau ;x_0 )
& = &(1+ \tau )^{-   i m } {\mathcal E}_{A}(x,\tau ;x_0 )+ \int_0^{\tau } \Big[ K_0(r,\tau  ;  m)+  2 i  m  K_1(r,\tau  ;  m)\Big] {\mathcal E}_{A}(x,r;x_0 )\,dr\,,\\
{\mathcal E}_{1}(x,\tau ;x_0 )
& = &
 \int_0^{\tau }   K_1(r,\tau  ; m)   {\mathcal E}_{A}(x,r;x_0 )\,dr
\end{eqnarray*}
solve  the Cauchy problem
\begin{eqnarray*}  
\cases{ 
\dsp \partial_\tau^2 {\mathcal E}_{i}(x,\tau ;x_0 )-  A(x,\partial_x) {\mathcal E}_{i}(x,\tau ;x_0 ) +\frac{ 2 i m}{ \tau +1 }\partial_\tau{\mathcal E}_{i}(x,\tau ;x_0 )  =0\,, \cr
{\mathcal E}_{i}(x,0;x_0 )=\delta _{i0}\delta  (x-x_0), \quad \partial_\tau{\mathcal E}_{i}(x,0;x_0 )= \delta_ {i1}\delta  (x-x_0)\,, \quad i=0,1\,.} 
\end{eqnarray*}
\end{theorem}

\subsection{Proof of Theorem~\ref{T5.4} for problem    with $\varphi _0 $. Case of $f=0$, $\varphi _1=0$}

Next, we will suppress  subscript $\varphi _0 $ of $v_{\varphi _0}(x,\tau ) $. It is convenient to split the solution 
$u(x,\tau )$ into two parts $u(x,\tau )=\widetilde u(x,\tau )+{\widetilde {\widetilde  u}} (x,\tau ) $, where
\begin{eqnarray}
\label{utilde}
\widetilde u(x,\tau )
& = &
(1+ \tau )^{-  i m } v(x,\tau )+ \int_0^{\tau }  K_0(r,\tau  ; m)  v(x,r)\,dr\,,\\
\label{u2tilde}
{\widetilde {\widetilde  u}} (x,\tau ) 
& = &
  2 i m  \int_0^{\tau }  K_1(r,\tau  ; m) v(x,r)\,dr\,.
\end{eqnarray}
It is easy to see that
$
u(x,0 )
= \varphi _0 (x  )
$
and
\begin{eqnarray*}
u_\tau (x,\tau )
& = &
\partial_\tau \left( (1+ \tau )^{-   i m } v(x,\tau ) \right) +\partial_\tau \int_0^{\tau } \left[ K_0(r,\tau  ; m)+  2 i m  K_1(r,\tau  ; m)\right] v(x,r)\,dr\\
& = &
-   i m   (1+ \tau )^{-   i m -1} v (x,\tau )   +   (1+ \tau )^{-   i m } \partial_\tau v (x,\tau )  \\
&  &
+ \left[ K_0(\tau ,\tau  ; m)+  2 i m  K_1(\tau ,\tau  ; m)\right] v (x,\tau ) \\
&  &
+\int_0^{\tau } \left[ \partial_\tau K_0(r,\tau  ; m)+  2 i m  \partial_\tau K_1(r,\tau  ; m)\right] v (x,r)\,dr\,.
\end{eqnarray*}
It follows from Lemma~\ref{L7.3} that
\[
\lim_{\tau \to 0} u_\tau (x,\tau )=
- \lim_{\tau \to 0}  i m   v(x,0)   
+\lim_{\tau \to 0} \left[ K_0(\tau ,\tau  ; m)+  2 i m  K_1(\tau ,\tau  ; m)\right] v(x,0 ) =0\,.
\]
According to Theorem~\ref{T5.4} the function ${\widetilde {\widetilde  u}} (x,\tau )  $ (\ref{u2tilde}) solves the equation, therefore it remains to verify that the function $\widetilde u(x,\tau ) $ (\ref{utilde}) solves the equation. For the first order derivative we obtain
\begin{eqnarray*}
\widetilde u_\tau (x,\tau )
& = &
\partial_\tau \left( (1+ \tau )^{-   i m } v(x,\tau ) \right) + \partial_\tau \int_0^{\tau }   K_0(r,\tau  ; m)  v(x,r)\,dr\\
& = &
-   i m   (1+ \tau )^{-  i m -1} v (x,\tau )   +   (1+ \tau )^{-   i m } \partial_\tau v (x,\tau )  \\
&  &
+    K_0(\tau ,\tau  ; m) v (x,\tau ) 
+\int_0^{\tau }  K_{0\,\tau }(r,\tau  ; m) v (x,r)\,dr\,.
\end{eqnarray*}
For the second order derivative we have
\begin{eqnarray*}
\widetilde u_{\tau \tau } (x,\tau ) 
& = &
   -   i m (-   i m -1 )  (1+ \tau )^{-   i m -2} v (x,\tau ) - 2 i m   (1+ \tau )^{-   i m -1} v _\tau (x,\tau )   +   (1+ \tau )^{-  i m }  v_{\tau \tau} (x,\tau  )  \\
&  &
+\partial_\tau  \Bigg\{  K_0(\tau ,\tau  ; m) v (x,\tau ) 
+\int_0^{\tau }   K_{0\,\tau } (r,\tau  ; m) v (x,r)\,dr\Bigg\}\,.
\end{eqnarray*}
Thus,  
\begin{eqnarray*}
\widetilde u_{\tau \tau } (x,\tau ) 
& = &
   -   i m (-   i m -1 )  (1+ \tau )^{-   i m -2} v (x,\tau ) - 2 i m   (1+ \tau )^{-   i m -1} v _\tau (x,\tau )   +   (1+ \tau )^{-   i m }  v_{\tau \tau} (x,\tau  )  \\
&  &
+ \Bigg[  K_{0\,r} (\tau ,\tau  ; m) v (x,\tau ) + 2K_{0\,\tau } (\tau ,\tau  ; m) v (x,\tau )+ K_0  (\tau ,\tau  ; m) v_\tau  (x,\tau )  \\
  &  &
 +   \int_0^{\tau }   K_{0\,\tau \tau  } (r,\tau  ; m) v (x,r)\,dr \Bigg] \,.
\end{eqnarray*}
Next we consider an application of the operator $A(x,\partial_x) $ to  the function  (\ref{utilde}): 
\begin{eqnarray*}
A(x,\partial_x)\widetilde u(x,\tau )
& = &
A(x,\partial_x)  (1+ \tau )^{-   i m } v(x,\tau )+ \int_0^{\tau }  K_0(r,\tau  ; m) A(x,\partial_x) v(x,r)\,dr  \\
& = &
  (1+ \tau )^{-   i m }A(x,\partial_x) v(x,\tau )+ B(x,\tau )  \,,
\end{eqnarray*}
where we have denoted 
\begin{eqnarray*}
B(x,\tau ) 
& := & 
\int_0^{\tau }  K_0(r,\tau  ; m)  v_{rr}(x,r)\,dr \,. 
\end{eqnarray*}
Integrating by parts and using $K_0(0,\tau  ; m)=0 $, we obtain
\begin{eqnarray*}
B(x,\tau )
& = & 
K_0(\tau ,\tau  ; m)  v_{r }(x,\tau )- \int_0^{\tau }  K_{0\,r} (r,\tau  ; m)  v_{r }(x,r)\,dr\\
&  = & 
K_0(\tau ,\tau  ; m)  v_{r }(x,\tau )-    K_{0\,r} (\tau ,\tau  ; m)  v (x,\tau )+ K_{0\,r} (0,\tau  ; m)  v (x,0)\\
&  &
+\int_0^{\tau }  K_{0\,r r} (r,\tau  ; m)  v (x,r)\,dr  \,. 
\end{eqnarray*}
Since $K_{0\,r} (0,\tau  ; m)=0$, 
it follows 
\begin{eqnarray*}
A(x,\partial_x)\widetilde u(x,\tau )
& = &
  (1+ \tau )^{-   i m }A(x,\partial_x) v(x,\tau )
+\Bigg[ K_0(\tau ,\tau  ; m)  v_{r }(x,\tau )-    K_{0\,r} (\tau ,\tau  ; m)  v (x,\tau )\\
&  &
+ K_{0\,r} (0,\tau  ; m)  v (x,0)+\int_0^{\tau }  K_{0\,r r} (r,\tau  ; m)  v (x,r)\,dr \Bigg]  \,. 
\end{eqnarray*}
Thus,
\begin{eqnarray*}
&  & 
\widetilde u_{\tau \tau }(x,\tau )-   A(x,\partial_x) \widetilde u(x,\tau ) +\frac{ 2 i m}{ \tau +1 }\widetilde u_\tau(x,\tau )  \\
& = & 
\Bigg\{    -   i m (-  i m -1 )  (1+ \tau )^{-   i m -2} v (x,\tau ) - 2 i m   (1+ \tau )^{-   i m -1} v _\tau (x,\tau )   +   (1+ \tau )^{-  i m }  v_{\tau \tau} (x,\tau  )  \\
&  &
+ \Bigg[  K_{0\,r} (\tau ,\tau  ; m) v (x,\tau ) + 2K_{0\,\tau } (\tau ,\tau  ; m) v (x,\tau )+ K_0  (\tau ,\tau  ; m) v_\tau  (x,\tau )  \\
  &  &
 +   \int_0^{\tau }   K_{0\,\tau \tau  } (r,\tau  ; m) v (x,r)\,dr \Bigg]\Bigg\}  
-    \Bigg\{  (1+ \tau )^{-   i m }A(x,\partial_x) v(x,\tau )\\
&  &
+\Bigg[ K_0(\tau ,\tau  ; m)  v_{r }(x,\tau )-    K_{0\,r} (\tau ,\tau  ; m)  v(x,\tau )+\int_0^{\tau }  K_{0\,r r} (r,\tau  ; m)  v (x,r)\,dr\Bigg]\Bigg\} \\
&  &
+\frac{ 2 i m}{ \tau +1 }\Bigg\{ -   i m   (1+ \tau )^{-  i m -1} v (x,\tau )   +   (1+ \tau )^{-   i m }   v_\tau (x,\tau )  \\
&  &
+ K_0(\tau ,\tau  ; m) v (x,\tau ) 
+ \int_0^{\tau }  K_{0\,\tau }(r,\tau  ; m) v (x,r)\,dr\Bigg\}  \,.
\end{eqnarray*}
By application of (\ref{K0eq})  
and the definition of the function $v=v(x,\tau ) $ 
we obtain
\begin{eqnarray*}
&  & 
\widetilde u_{\tau \tau }(x,\tau )-   A(x,\partial_x) \widetilde u(x,\tau ) +\frac{ 2 i m}{ \tau +1 }\widetilde u_\tau(x,\tau )  \\
& = & 
\Bigg[    -   i m (-   i m -1 )  (1+ \tau )^{-   i m -2}    
+2 K_{0\,r} (\tau ,\tau  ; m)  + 2K_{0\,\tau } (\tau ,\tau  ; m)
    \\
&  &
+\frac{ 2 i m}{ \tau +1 }\Bigg\{ -  i m   (1+ \tau )^{-   i m -1}     
+ K_0(\tau ,\tau  ; m)\Bigg\} \Bigg] v (x,\tau ) \\ 
& =  & 
  m ( m +im )  (1+ \tau )^{-   i m -2} v (x,\tau )\\
&  &
+ \Bigg[ 2 K_{0\,r} (\tau ,\tau  ; m)v (x,\tau )  + 2K_{0\,\tau } (\tau ,\tau  ; m) +  \frac{ 2 i m}{ \tau +1 }   K_0(\tau ,\tau  ; m)\Bigg] v (x,\tau )\,.
\end{eqnarray*}
Then we apply the last equation of Proposition~\ref{MN_P2.9} and obtain 
\[  
  m ( m +im )  (1+ \tau )^{-   i m -2} 
+\Bigg[ 2 K_{0\,r} (\tau ,\tau  ; m)  + 2K_{0\,\tau } (\tau ,\tau  ; m) +  \frac{ 2 i m}{ \tau +1 }   K_0(\tau ,\tau  ; m)\Bigg]   =0 \,.
\]
Thus, this case of Theorem~\ref{T5.4} is proved. 
\qed

\subsection{Proof of Theorem~\ref{T5.4} for problem  with $\varphi _1 $. Case of $f=0$, $\varphi _0=0$}

 Next, we will suppress  subscript $\varphi _1 $ of $v_{\varphi _1}(x,\tau ) $. It is evident that 
$
u(x,0 )=0 
$. 
For the derivative $ \partial_\tau u(x,\tau )$  we have 
\begin{eqnarray}
\partial_\tau u(x,\tau )
& = &
\partial_\tau  \int_0^{\tau }   K_1(r,\tau  ; m)   v(x,r)\,dr 
\label{utauder}
=
K_1(\tau ,\tau  ; m)   v(x,\tau ) +\int_0^{\tau }   K_{1\,\tau }(r,\tau  ; m)   v(x,r)\,dr\,.
\end{eqnarray}
In particular,
\begin{eqnarray*} 
\lim_{\tau \to 0} \partial_\tau u(x,\tau )
& = &
\lim_{\tau \to 0} 
 K_1(\tau ,\tau  ; m)   v(x,\tau ) = \varphi_1   (x )\,,
\end{eqnarray*} 
where we have used 
\begin{eqnarray*} 
K_1(\tau ,\tau  ; m)
& = &
 { 2^{2 i  m}}\left((\tau +2)^2-\tau ^2\right)^{-  i m}\,.
\end{eqnarray*} 
Then we use (\ref{utauder}) to find the second order derivative 
\[
\partial_\tau^2 u(x,\tau )
=
   K_{1\,r}(\tau ,\tau  ; m)   v (x,\tau )  +  2 K_{1\,\tau }(\tau ,\tau  ; m)   v (x,\tau )+  K_{1}(\tau ,\tau  ; m)   v_\tau  (x,\tau )
+  \int_0^{\tau }   K_{1\,\tau \tau  }(r,\tau  ; m)   v_{ }(x,r)\,dr \,.
\]
Using the definition of the function $v (x,r) $   and the  integration by parts twice, we derive 
\begin{eqnarray*}
 A(x,\partial_x)u(x,\tau )
& = &
\int_0^{\tau }   K_1(r,\tau  ; m)   A(x,\partial_x)v (x,r)\,dr \\
& = &
\int_0^{\tau }   K_1(r,\tau  ; m)    v_{r r} (x,r)\,dr \\
& = &
 K_1(\tau ,\tau  ; m)    v_{r } (x,\tau )- \Bigg\{  K_{1\,r}(\tau ,\tau  ; m)    v  (x,\tau )- K_{1\,r}(0,\tau  ; m)    v  (x,0)\\
&  &
- \int_0^{\tau }   K_{1\,r r}(r,\tau  ; m)    v  (x,r)\,dr \Bigg\}\,.
\end{eqnarray*}
Since $
K_{1\,r}(0,\tau  ; m)=0
$,
we obtain
\[
A(x,\partial_x)u(x,\tau )=
 K_1(\tau ,\tau  ; m)    v_{r } (x,\tau )-   K_{1\,r}(\tau ,\tau  ; m)    v  (x,\tau ) 
+ \int_0^{\tau }   K_{1\,r r}(r,\tau  ; m)    v  (x,r)\,dr \,. 
\]
Consequently, according to (\ref{K1eq}) of  Theorem~\ref{MNT2.12} and Proposition~\ref{MN_P2.9}  since $ v_r  (x,\tau )= v_\tau  (x,\tau ) $, we obtain 
\begin{eqnarray*}
&  & 
 u_{\tau \tau }(x,\tau )-   A(x,\partial_x)u(x,\tau )+\frac{ 2 i m}{ \tau +1 }u_\tau (x,\tau )\\
& = &
  \Bigg\{   K_{1\,r}(\tau ,\tau  ; m)   v (x,\tau ) +  K_{1}(\tau ,\tau  ; m)   v_\tau  (x,\tau )+   2K_{1\,\tau }(\tau ,\tau  ; m)   v (x,\tau ) 
+  \int_0^{\tau }   K_{1\,\tau \tau  }(r,\tau  ; m)   v_{ }(x,r)\,dr\Bigg\}\\
&  &
-     \Bigg\{ K_1(\tau ,\tau  ; m)    v_{r } (x,\tau )-   K_{1\,r}(\tau ,\tau  ; m)    v  (x,\tau )+ \int_0^{\tau }   K_{1\,r r}(r,\tau  ; m)    v  (x,r)\,dr\Bigg\}\\
&  &
+  \frac{ 2 i m}{ \tau +1 }\Bigg\{ K_1(\tau ,\tau  ; m)   v(x,\tau ) +\int_0^{\tau }   K_{1\,\tau }(r,\tau  ; m)   v(x,r)\,dr\Bigg\}  \\
& = &
  \Bigg\{ 2  K_{1\,r}(\tau ,\tau  ; m)   +   2K_{1\,\tau }(\tau ,\tau  ; m)    
+  \frac{ 2 i m}{ \tau +1 }  K_1(\tau ,\tau  ; m) \Bigg\}  v (x,\tau )=0\,.
\end{eqnarray*}
Thus, this case of Theorem~\ref{T5.4} is proved.  \qed

\subsection{Proof of Theorem~\ref{T5.4} for problem  with $f $. Case of  $\varphi _0=\varphi _1=0$}

 Next, we will suppress  subscript $f $ of  $v_{f}(x,r;b) $. We write  $v (x,r;b) $ for  $v_{f}(x,r;b) $ and calculate the derivative
\begin{eqnarray*}
\partial_\tau  u(x,\tau )
& = &
 \int_0^{\tau } \,db \,\partial_\tau\int_0^{\tau -b}  E(r,\tau;b  ; m)   v(x,r;b)\,dr \\
& = &
 \int_0^{\tau }     E(\tau -b,\tau;b  ; m)   v(x,\tau -b;b)\,db  
 +  \int_0^{\tau } \,db \int_0^{\tau -b}   E_\tau (r,\tau;b  ; m)   v(x,r;b)\,dr\,.
\end{eqnarray*}
It follows that the function takes required initial data as well as 
\begin{eqnarray*}
\partial_\tau^2  u(x,\tau )
& = &
     E(0,\tau;\tau   ; m)    f (x, \tau ) 
+ \int_0^{\tau }     E_r(\tau -b,\tau;b  ; m)   v (x,\tau -b;b)\,db\\
&  &
+ \int_0^{\tau }     E (\tau -b,\tau;b  ; m)   v_\tau (x,\tau -b;b)\,db +
2 \int_0^{\tau }       E_\tau(\tau -b,\tau;b  ; m)    v (x,\tau -b;b)\,db \\
&  &
+   \int_0^{\tau } \,db \int_0^{\tau -b}  \partial_\tau^2 E(r,\tau;b  ; m)   v (x,r;b)\,dr  \,.
\end{eqnarray*}
Then   using the definition of the function $v=v(x,r)$ and the integration by parts twice, we derive 
\begin{eqnarray*}
A(x,\partial_x) u(x,\tau )
& = &
\int_0^{\tau } \,db\int_0^{\tau -b}  E(r,\tau; b  ; m)  A(x,\partial_x)   v (x,r;b)\,dr \\
& = &
\int_0^{\tau } \,db\int_0^{\tau -b}  E(r,\tau; b  ; m)  \partial_r^2    v (x,r;b)\,dr \\
& = &
 \int_0^{\tau } \,db\Bigg[  E(\tau -b,\tau; b  ; m)        v_r  (x,\tau -b;b)
- E(0,\tau; b  ; m)    v_r  (x,0;b)
\\
&  &
- \int_0^{\tau -b}  \partial_r E(r,\tau; b  ; m)      v_r (x,r;b)\,dr \Bigg] \\
& = &
 \int_0^{\tau } \,db\Bigg[  E(\tau -b,\tau; b  ; m)        v_r  (x,\tau -b;b)
- \int_0^{\tau -b}  \partial_r E(r,\tau; b  ; m)      v_r (x,r;b)\,dr \Bigg]  \,.
\end{eqnarray*}
Hence
\begin{eqnarray*}
&  & 
 u_{\tau \tau }(x,\tau )-   A(x,\partial_x)u(x,\tau )+\frac{ 2 i m}{ \tau +1 }u_\tau(x,\tau ) \\
& = &
\Bigg[     E(0,\tau;\tau   ; m)    f (x, \tau ) 
+ \int_0^{\tau }     E_r(\tau -b,\tau;b  ; m)   v (x,\tau -b;b)\,db\\
&  &
+ \int_0^{\tau }     E (\tau -b,\tau;b  ; m)   v_\tau (x,\tau -b;b)\,db +
2 \int_0^{\tau }       E_\tau(\tau -b,\tau;b  ; m)    v (x,\tau -b;b)\,db \\
&  &
+   \int_0^{\tau } \,db \int_0^{\tau -b}  \partial_\tau^2 E(r,\tau;b  ; m)   v (x,r;b)\,dr \Bigg] \\
&  &
-\Bigg[   \int_0^{\tau } \,db\Bigg\{  E(\tau -b,\tau; b  ; m)        v_r  (x,\tau -b;b)
- \int_0^{\tau -b}  \partial_r E(r,\tau; b  ; m)      v_r (x,r;b)\,dr \Bigg\}\Bigg] \\
&  &
+\frac{ 2 i m}{ \tau +1 }\Bigg[  \int_0^{\tau }     E(\tau -b,\tau;b  ; m)   v(x,\tau -b;b)\,db  
 +  \int_0^{\tau } \,db \int_0^{\tau -b}   E_\tau (r,\tau;b  ; m)   v(x,r;b)\,dr \Bigg]  \,.
 \end{eqnarray*}
 On the other hand, the integration by parts leads to
\begin{eqnarray*}
&  &
 \int_0^{\tau } \,db \int_0^{\tau -b}   E_r(r,\tau; b  ; m)      v_r (x,r;b)\,dr\\
  &  = &
\int_0^{\tau } \Bigg\{   E_r (\tau -b,\tau; b  ; m)      v (x,\tau -b;b)
-   E_r(0,\tau; b  ; m)      v  (x,0;b)- \int_0^{\tau -b}   E_{rr}(r,\tau; b  ; m)      v  (x,r;b)\,dr \Bigg\}  \,db\,.
 \end{eqnarray*} 
 Then we use $
E_r(0,\tau; b  ; m)
= 0  $. 
 Consequently,
 \begin{eqnarray*}
&  &
 \int_0^{\tau } \,db \int_0^{\tau -b}  \partial_r E(r,\tau; b  ; m)      v_r (x,r;b)\,dr\\
  &  = &
\int_0^{\tau } \Bigg\{  \partial_r E(\tau -b,\tau; b  ; m)      v (x,\tau -b;b)
- \partial_r E(0,\tau; b  ; m)      f  (x, b)- \int_0^{\tau -b}  \partial_{rr} E(r,\tau; b  ; m)      v  (x,r;b)\,dr \Bigg\} \,db\\
  &  = &
\int_0^{\tau }     E_r(\tau -b,\tau; b  ; m)      v (x,\tau -b;b) \,db
- \int_0^{\tau }  \,db\int_0^{\tau -b}    E_{rr}(r,\tau; b  ; m)      v  (x,r;b)\,dr \,.
 \end{eqnarray*} 
 Thus,
\begin{eqnarray*}
&  & 
u_{\tau \tau }(x,\tau )-   A(x,\partial_x)u(x,\tau )+\frac{ 2 i m}{ \tau +1 }u_\tau (x,\tau )\\
& = &
\Bigg[     E(0,\tau;\tau   ; m)    f (x, \tau ) 
+ \int_0^{\tau }     E_r(\tau -b,\tau;b  ; m)   v (x,\tau -b;b)\,db\\
&  &
+ \int_0^{\tau }     E (\tau -b,\tau;b  ; m)   v_\tau (x,\tau -b;b)\,db +
2 \int_0^{\tau }       E_\tau(\tau -b,\tau;b  ; m)    v (x,\tau -b;b)\,db \\
&  &
+   \int_0^{\tau } \,db \int_0^{\tau -b}  \partial_\tau^2 E(r,\tau;b  ; m)   v (x,r;b)\,dr \Bigg] 
-    \int_0^{\tau }    E(\tau -b,\tau; b  ; m)        v_r  (x,\tau -b;b)\,db
\\
&  &
+ \Bigg\{\int_0^{\tau }     E_r(\tau -b,\tau; b  ; m)      v (x,\tau -b;b) \,db
- \int_0^{\tau }  \,db\int_0^{\tau -b}    E_{rr}(r,\tau; b  ; m)      v  (x,r;b)\,dr\Bigg\} \\
&  &
+\frac{ 2 i m}{ \tau +1 }\Bigg[  \int_0^{\tau }     E(\tau -b,\tau;b  ; m)   v(x,\tau -b;b)\,db  
 +  \int_0^{\tau } \,db \int_0^{\tau -b}   E_\tau (r,\tau;b  ; m)   v(x,r;b)\,dr \Bigg].
 \end{eqnarray*} 
 The double integrals of the last equation can be unified as follows
 \begin{eqnarray*}
&  &
 \int_0^{\tau } \,db \int_0^{\tau -b}  \partial_\tau^2 E(r,\tau;b  ; m)   v (x,r;b)\,dr 
- \int_0^{\tau }  \,db\int_0^{\tau -b}    E_{rr}(r,\tau; b  ; m)      v  (x,r;b)\,dr  \\
&  &
 +\frac{ 2 i m}{ \tau +1 } \int_0^{\tau } \,db \int_0^{\tau -b}   E_\tau (r,\tau;b  ; m)   v(x,r;b)\,dr \\
& = &
 \int_0^{\tau } \,db \int_0^{\tau -b} \Bigg\{   E_{\tau \tau } (r,\tau;b  ; m)
-     E_{rr}(r,\tau; b  ; m)       
 +\frac{ 2 i m}{ \tau +1 }     E_\tau (r,\tau;b  ; m) \Bigg\}  v(x,r;b) \,dr=0 \,.
\end{eqnarray*} 
Hence, according to (\ref{Eeq}) we derive
\begin{eqnarray*}
&  & 
 u_{\tau \tau }(x,\tau )-   A(x,\partial_x)u(x,\tau )+\frac{ 2 i m}{ \tau +1 }u_\tau(x,\tau )\\
& = &
 E(0,\tau;\tau   ; m)    f (x, \tau ) 
+ \int_0^{\tau }     E_r(\tau -b,\tau;b  ; m)   v (x,\tau -b;b)\,db\\
&  &
+ \int_0^{\tau }     E (\tau -b,\tau;b  ; m)   v_\tau (x,\tau -b;b)\,db +
2 \int_0^{\tau }       E_\tau(\tau -b,\tau;b  ; m)    v (x,\tau -b;b)\,db \\
&  &
-    \int_0^{\tau }    E(\tau -b,\tau; b  ; m)        v_r  (x,\tau -b;b)\,db
+  \int_0^{\tau }     E_r(\tau -b,\tau; b  ; m)      v (x,\tau -b;b) \,db \\
&  &
+\frac{ 2 i m}{ \tau +1 }   \int_0^{\tau }     E(\tau -b,\tau;b  ; m)   v(x,\tau -b;b)\,db   \,.
 \end{eqnarray*} 
Next we use $  v_\tau (x,\tau -b;b)=  v_r  (x,\tau -b;b)$ in 
\begin{eqnarray*}
&  &
\int_0^{\tau }     E (\tau -b,\tau;b  ; m)   v_\tau (x,\tau -b;b)\,db-    \int_0^{\tau }    E(\tau -b,\tau; b  ; m)        v_r  (x,\tau -b;b)\,db
=0  \,.
\end{eqnarray*} 
Hence
\begin{eqnarray*}
&  & 
 u_{\tau \tau }(x,\tau )-   A(x,\partial_x)u(x,\tau )+\frac{ 2 i m}{ \tau +1 }u_\tau (x,\tau )\\
& = &
 E(0,\tau;\tau   ; m)    f (x, \tau ) \\
&  &
+ \int_0^{\tau }     E_r(\tau -b,\tau;b  ; m)   v (x,\tau -b;b)\,db+
2 \int_0^{\tau }       E_\tau(\tau -b,\tau;b  ; m)    v (x,\tau -b;b)\,db 
\\
&  &
+  \int_0^{\tau }     E_r(\tau -b,\tau; b  ; m)      v (x,\tau -b;b) \,db 
+\frac{ 2 i m}{ \tau +1 }   \int_0^{\tau }     E(\tau -b,\tau;b  ; m)   v(x,\tau -b;b)\,db  \\
& = &
 E(0,\tau;\tau   ; m)    f (x, \tau ) \\
 &  &
+ \int_0^{\tau }    \Bigg[  2E_r(\tau -b,\tau;b  ; m) 
+   2       E_\tau(\tau -b,\tau;b  ; m)  
+\frac{ 2 i m}{ \tau +1 }      E(\tau -b,\tau;b  ; m) \Bigg]  v(x,\tau -b;b)\,db   \,.
\end{eqnarray*}
Then we apply Proposition~\ref{MN_P2.9}
and  obtain 
\begin{eqnarray*} 
 u_{\tau \tau }(x, \tau )-   A(x,\partial_x)u(x, \tau )+\frac{ 2 i m}{ \tau +1 }u_\tau (x, \tau ) 
& = &  f (x, \tau ) \,.
\end{eqnarray*}
This case of Theorem~\ref{T5.4} is proved.  Theorem~\ref{T5.4} is proved. \qed

\end{document}